\documentclass[fleqn,10pt]{wlscirep}
\usepackage[utf8]{inputenc}
\usepackage[T1]{fontenc}
\usepackage{amsmath}
\usepackage{enumitem}
\usepackage{tcolorbox}
\usepackage{graphicx}
\usepackage{amssymb}
\usepackage{epstopdf}
\usepackage{url}
\usepackage{subfigure}
\usepackage{color}

\usepackage{mathtools}
\usepackage{color, colortbl}
\usepackage{fancyhdr}
\usepackage{ctable}
\definecolor{Gray}{gray}{0.9}
\usepackage{enumitem}
\usepackage{graphicx}
\usepackage{caption}
\usepackage{color, colortbl}
\usepackage{array}
\usepackage{booktabs} 
\usepackage{makecell}
\usepackage{cellspace}
\usepackage{tabularx}
\usepackage{multirow}
\usepackage{amsmath}
\usepackage{listings}
\usepackage{nameref}

\usepackage{xr}

\usepackage{verbatim}

\lstdefinelanguage{json}{
    basicstyle=\ttfamily\small,
    showstringspaces=false,
    breaklines=true,
    frame=single,
    morestring=[b]",
    moredelim=[s][\bfseries\color{black}]{:}{\ },
}

\newcommand\setcurrentname[1]{\def\@currentlabelname{#1}}
\makeatother

\title{The Potential Impact of Disruptive AI Innovations on U.S. Occupations}

\author[1, +]{Munjung Kim}
\author[2, 3]{Marios Constantinides}
\author[2, 4]{Sanja \v{S}\'{c}epanovi\'{c}}
\author[1]{Yong-Yeol Ahn}
\author[2,5,*]{Daniele Quercia}
\affil[1]{Indiana University Bloomington, United States}
\affil[2]{Nokia Bell Labs, Cambridge, United Kingdom}
\affil[3]{CYENS Centre of Excellence, Nicosia, Cyprus}
\affil[4]{The Uehiro Oxford Institute, Oxford, United Kingdom}
\affil[5]{Politecnico di Torino, Turin, Italy}

\affil[*]{quercia@cantab.net}
\affil[+]{Work done at Nokia Bell Labs}

\newcolumntype{R}[1]{>{\raggedleft\arraybackslash}p{#1}}

\newcommand{\revision}[1]{\textcolor{blue}{#1}}

\renewcommand{\revision}[1]{#1}

\begin{abstract}

The rapid rise of AI is poised to disrupt the labor market. However, AI is not a monolith; its impact depends on both the nature of the innovation and the jobs it affects. While computational approaches are emerging, there is no consensus on how to systematically measure an innovation's disruptive potential. Here, we calculate the disruption index of 3,237 U.S. AI patents (2015-2022) and link them to job tasks to distinguish between ``consolidating'' AI innovations that reinforce existing structures and ``disruptive'' AI innovations that alter them. Our analysis reveals that consolidating AI primarily targets physical, routine, and solo tasks, common in manufacturing and construction in the Midwest and central states. By contrast, disruptive AI affects unpredictable and mental tasks, particularly in coastal science and technology sectors. Surprisingly, we also find that disruptive AI disproportionately affects areas already facing skilled labor shortages, suggesting disruptive AI technologies may accelerate change where workers are scarce rather than replacing a surplus. Ultimately, consolidating AI appears to extend current automation trends, while disruptive AI is set to transform complex mental work, with a notable exception for collaborative tasks.

\end{abstract}

\RequirePackage[explicit]{titlesec}
\usepackage{hyperref}

\titleformat{name=\section,numberless}
  {\large\sffamily\bfseries}
  {}
  {0em}
  {\leavevmode\MakeLinkTarget[section]{}\ignorespaces#1}
  []  

\ExplSyntaxOn\makeatletter
\def\ttl@straight@i#1[#2]#3{%
  \tl_if_empty:nTF {#2}
   {\NR@gettitle{#3}}
   {\NR@gettitle{#2}}
  \gdef\ttl@savemark{\csname#1mark\endcsname{#3}}%
  \let\ttl@savewrite\@empty
  \def\ttl@savetitle{#3}%
  \gdef\thetitle{\csname the#1\endcsname}%
  \if@noskipsec \leavevmode \fi
  \par
  \ttl@labelling{#1}{#2}%
  \ttl@startargs\ttl@straight@ii{#1}{#3}}

\ExplSyntaxOff\makeatother

\begin{document}
\setcounter{secnumdepth}{0}
\flushbottom
\maketitle

\lstdefinelanguage{json}{
    basicstyle=\ttfamily\small,
    showstringspaces=false,
    breaklines=true,
    frame=single,
    morestring=[b]",
    moredelim=[s][\bfseries\color{black}]{:}{\ },
}

\thispagestyle{empty}

\section*{Introduction}

AI is changing how people work. Unlike earlier technologies, AI systems can learn and adapt to perform tasks. They are already taking on jobs that once required human intelligence, or creativity, raising serious concerns regarding its potential to drastically disrupt labor markets and the current economic order. For instance, studies show that AI tools can boost productivity in writing tasks, but also reduce opportunities for freelance work\cite{demirci2023ai, liu2023generate, noy2023experimental, agrawal2022chatgpt, gmyrek2023genai}. AI is already being adopted in many areas, including customer service, software development, and healthcare\cite{cappelli2024will, morandini2023impact}. Given the enormous potential and uncertainty around AI's impact on jobs, understanding how AI can reshape jobs is essential because such understanding may allow us to identify jobs and tasks that are at risks and how prepare the workforce to adapt accordingly.

Previous research on technological innovation suggests that there may be two distinct types of impact from disruptive technologies: \emph{consolidating} innovations improve existing tools or processes, while \emph{disruptive} innovations change how work is done and can make existing skills or systems obsolete\cite{tushman1986technological, baker2012technology,hci_disruption}.

\begin{figure}
    \centering
    \includegraphics[width=0.95\linewidth]{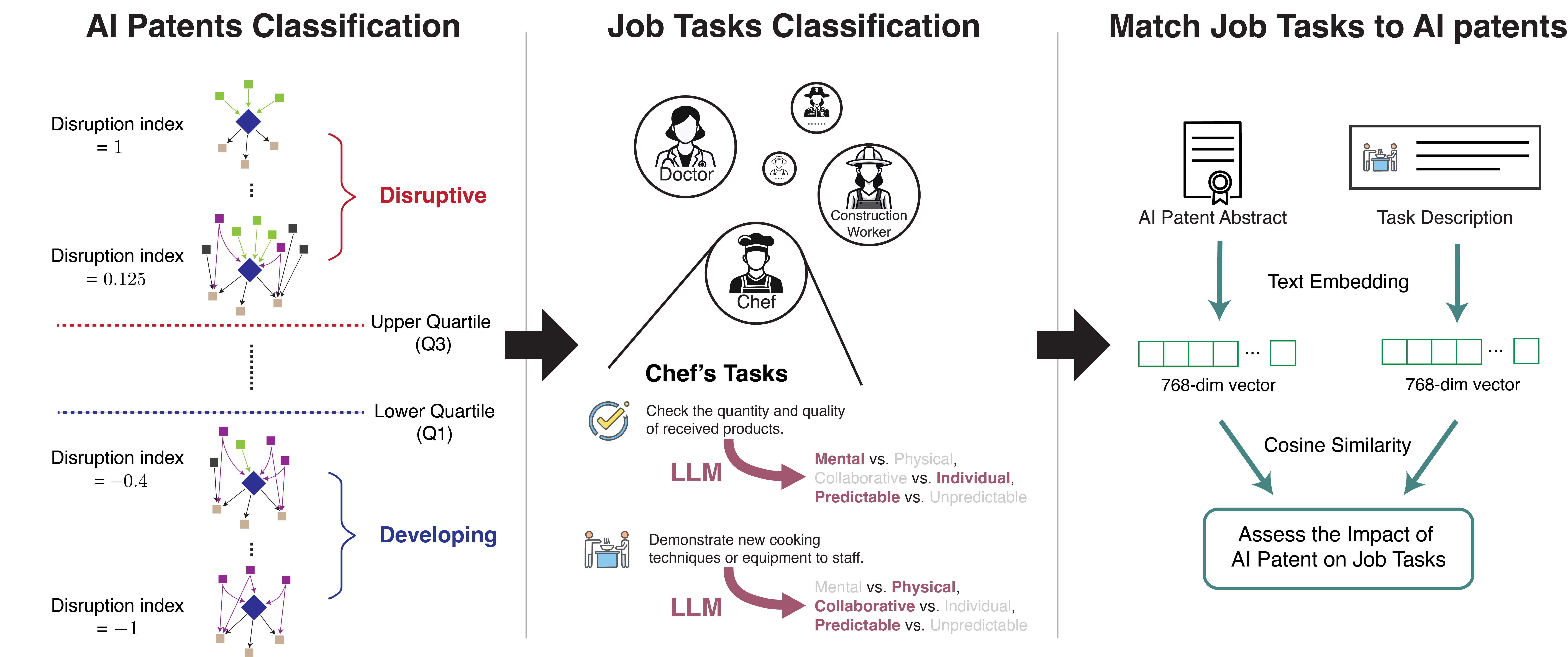}
    \caption{\textbf{Three steps to link AI patents to job tasks.} 
(Left) Step 1: We analyzed 3,237 AI patents from the U.S. Patent and Trademark Office (USPTO) and labeled each as disruptive or consolidating using the Disruption Index. Patents in the top 25\% were disruptive; those in the bottom 25\% were consolidating. 
(Middle) Step 2: We grouped 1,866 job tasks from the O*NET database using three dimensions: mental or physical, collaborative or individual, and predictable or unpredictable. 
(Right) Step 3: We matched tasks to patents using a language model (SBERT) that measured text similarity. A task was considered linked to a patent, if the similarity score passed a validated threshold.}
    \label{fig:enter-label}
\end{figure}

Building on this idea, we hypothesize that consolidating and disruptive AI technologies affect jobs differently and use the disruptiveness of AI patents to characterize related tasks, jobs, and industries. Specifically, we analyzed 3,237 AI patents from the U.S. Patent and Trademark Office (USPTO) and classify them. Using the Disruption Index~\cite{funk2017dynamic} (a bibliometric measure of patent citation patterns), we classified each patent as disruptive or consolidating (Figure~\ref{fig:enter-label} sketches the procedure and the  Methods section details it).

The index quantifies the extent to which a patent shifts the direction of subsequent innovation: patents with high disruption scores tend to break with established trajectories, while those with low scores reinforce existing ones~\cite{funk2017dynamic,wu2019large}. Patents in the top quartile of the index were labeled disruptive; those in the bottom quartile were labeled consolidating (Tables~\ref{tab:reppat}--\ref{tab:patent_disruption_75_80}). Next, we categorized 1,866 tasks from the O*NET database based on three dimensions: mental or physical, collaborative or individual, and predictable or unpredictable. These categories come from a review of past research on technology and work\cite{jesuthasan2021work, autor2013growth, autor2015untangling, deming2017growing}. We used GPT-4o to label tasks and validated the results with 24 human annotators (see \nameref{sec:methods}). Finally, similar to recent work\cite{septiandri2023impact, webb2019impact}, we matched each task to AI patents by comparing the text of task descriptions and patent abstracts. If the similarity score passed a validated threshold, the task was considered impacted by that patent.

\section*{Results}

\subsection*{Disruptive AI Affects Mental and Unpredictable Tasks but Not Collaborative Ones}

We began by examining how different types of AI patents relate to job tasks with specific characteristics. We hypothesized that disruptive AI patents would more often affect complex tasks, while consolidating AI patents would be linked to routine, predictable tasks.

To test this, we built a null model. We randomly assigned our task characteristics (mental \emph{vs.} physical, collaborative \emph{vs.} individual, and predictable \emph{vs.} unpredictable) while keeping their overall distribution fixed. We then calculated the share of tasks with each characteristic that were linked to disruptive or consolidating AI, or not linked to any AI. This process was repeated many times to create a baseline distribution. We compared observed proportions to this baseline using $z$-scores, which show how much the actual data differs from random chance.

\begin{figure}[t!]
    \centering
    \includegraphics[width=0.9\linewidth]{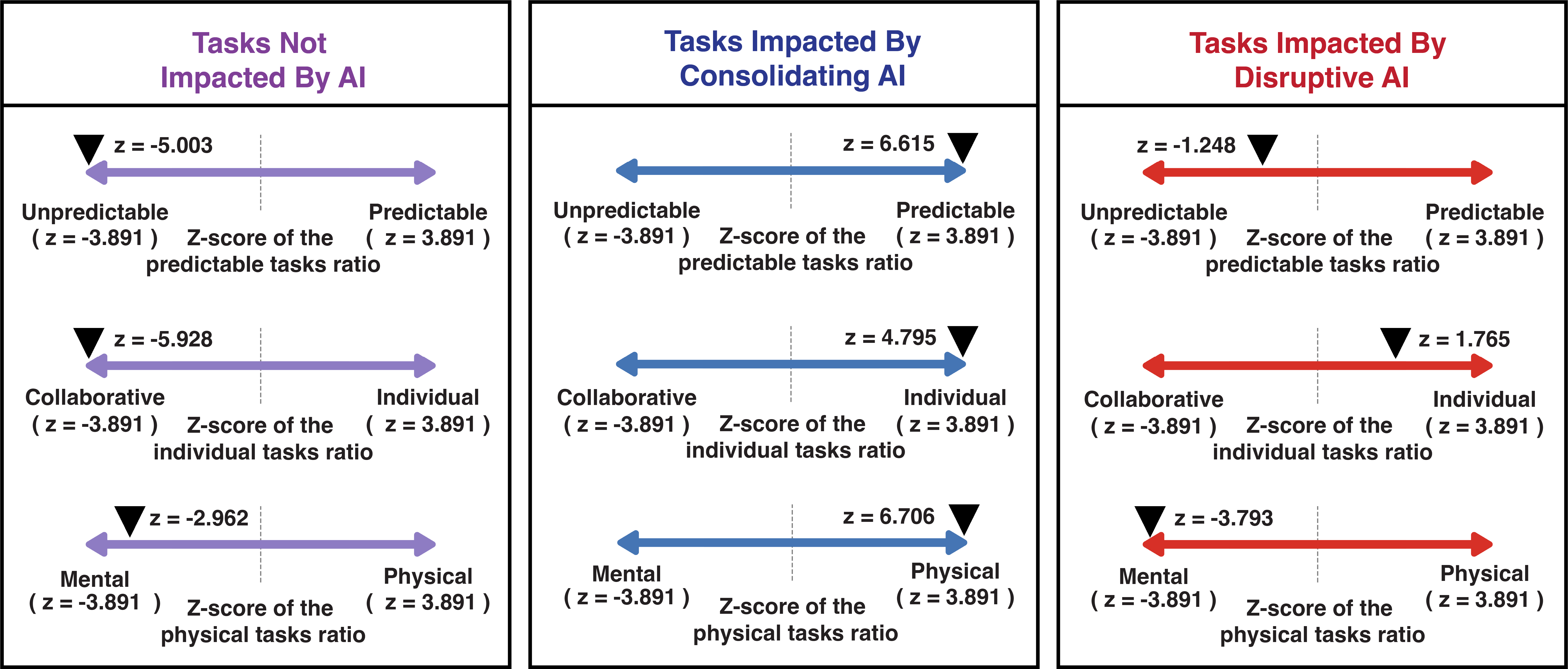}
    \caption{ \textbf{Task characteristics by AI patent type.} Each arrow shows how far the observed share of tasks with a certain characteristic deviates from what would be expected under random assignment. Black triangles show the actual $z$-scores. Positive values indicate a higher-than-expected presence of a trait (e.g., predictable, individual, or physical), while negative values suggest the opposite. Tasks \emph{not} impacted by AI tend to be unpredictable, collaborative, and mental. Tasks linked to consolidating AI are more likely to be predictable, individual, and physical. Disruptive AI targets mental and unpredictable tasks, but not collaborative ones.}
    \label{fig:tasks_characteristics}
\end{figure}

Figure~\ref{fig:tasks_characteristics} shows the results. Tasks linked to disruptive AI patents were more likely to be mental ($z = 3.793$) and unpredictable ($z = 1.248$). However, they remained individual rather than collaborative ($z = 1.765$). In contrast, tasks linked to consolidating AI were strongly associated with predictability ($z = 6.615$), physical effort ($z = 6.706$), and individual work ($z = 4.795$). Finally, tasks not affected by AI tended to be mental, collaborative, and unpredictable. While this might suggest they are safe from AI, a closer look shows that disruptive AI is already targeting mental and unpredictable work. Collaborative tasks, however, remain largely untouched. This aligns with recent arguments that current AI tools focus more on replacing solo tasks than on supporting teamwork~\cite{autor2024applying,Constantinides2025AI}.

\subsection*{Disruptive AI Affects Service Industries; Consolidating AI Affects Production}

We then studied how each industry is affected by disruptive and consolidating AI. We compared each industry's share of tasks impacted by disruptive or consolidating AI with its share of tasks impacted by general AI. This shows whether a type of AI innovation (e.g., disruptive) clusters in certain industries more than AI exposure overall. To do this, we linked tasks to jobs and assigned each job to its industry. We then calculated what proportion each industry represents among all tasks affected by disruptive AI, consolidating AI, and general AI.

\begin{figure}[t!]
    \centering
    \includegraphics[width=0.95\linewidth]{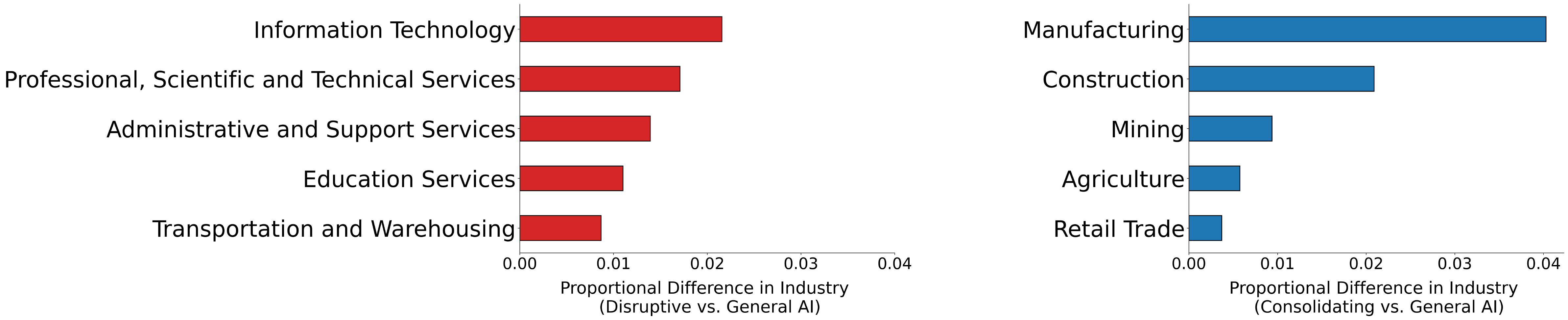}
    \caption{\textbf{Top five industry sectors most affected by each type of AI.} Bars show the difference in task impact between disruptive or consolidating AI and all AI patents. Positive values mean a sector is more affected by that type of AI than the overall average.}
    \label{fig:bubble_industry}
\end{figure}

Figure~\ref{fig:bubble_industry} shows that disruptive AI affects service-oriented sectors the most (Figure~\ref{fig:all_industry} has the results for all sectors). These include information technology (2.16\% above average), professional, scientific, and technical services (1.70\%), administrative and support services (1.39\%), education (1.10\%), and transportation (0.87\%). In contrast, consolidating AI is more common in production-heavy industries. These include manufacturing (4.03\% above average), construction (2.09\%), mining (0.94\%), agriculture (0.58\%), and retail trade (0.37\%).

\subsection*{Disruptive AI is Concentrated on the Coasts}

We looked at where disruptive and consolidating AI patents are developed. We compared U.S. states by the share of AI patents that impact job tasks and are classified as either disruptive or consolidating.

\begin{figure}[t]
    \centering
    \includegraphics[width=0.9\linewidth]{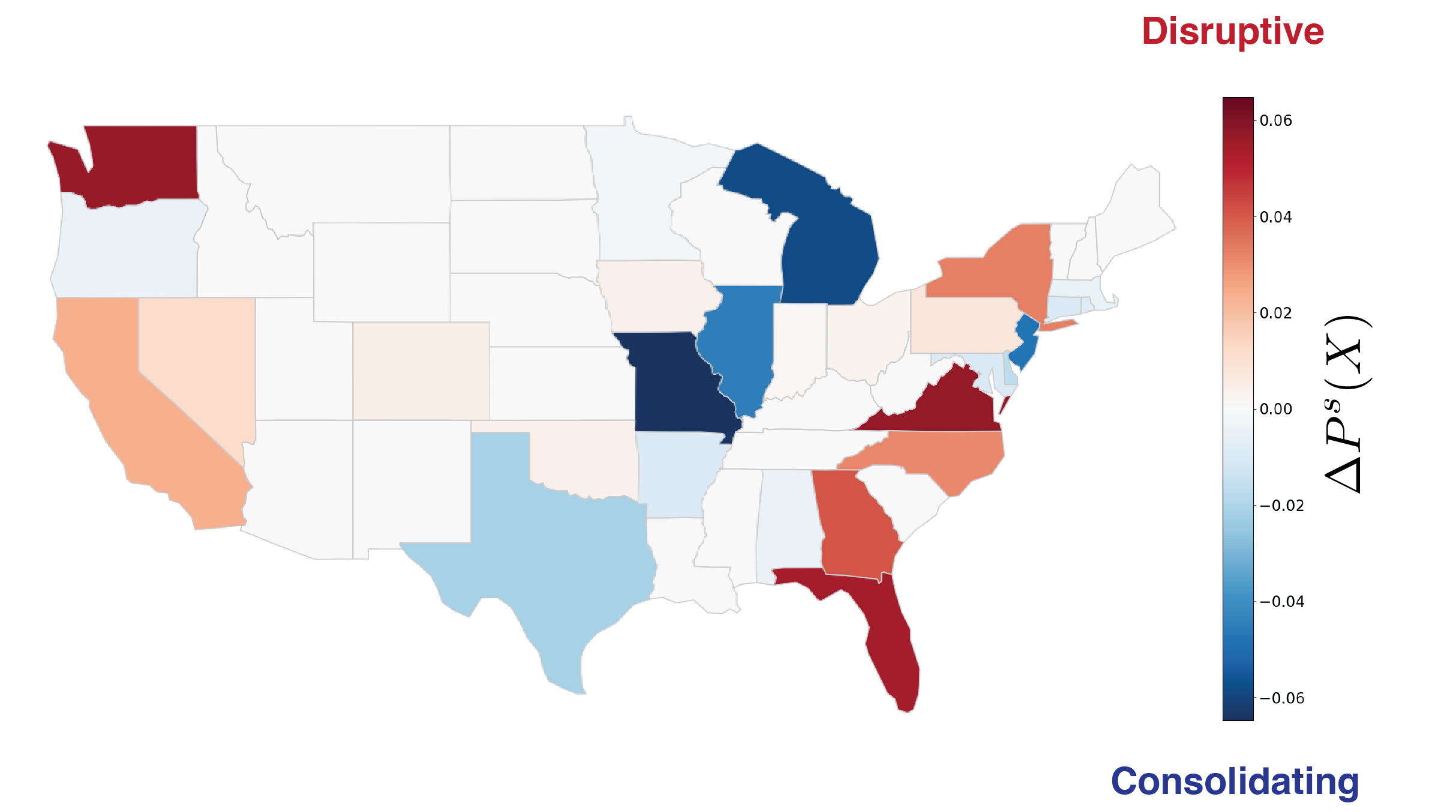}
    \caption{ \textbf{Regional patterns in AI development.} Red indicates states where disruptive AI patents are more prevalent; blue indicates a focus on consolidating AI. Color intensity reflects the difference in patent share compared to general AI. A value of 0.01 means that, in that state, disruptive AI patents affect 1\% more job tasks than consolidating AI patents.}
    \label{fig:agglomeration_State}
\end{figure}

Figure~\ref{fig:agglomeration_State} shows the geographic distribution. Red states have a higher share of disruptive patents. These include California, Washington, Florida, Virginia, and New York. Blue states, such as those in the Midwest, have more consolidating patents. This geographic divide suggests that disruptive AI, which affects complex, mental work, is being developed mainly in coastal innovation hubs. Meanwhile, consolidating AI, which ties to physical and routine tasks, is more common in central U.S. regions.

\revision{\subsection*{Disruptive AI Impacts Sectors Experiencing Labor Shortages}\label{sec:vacancy}}

\noindent We then investigated the association between job vacancy rates of each industry in 2022 and the extent to which each industry is affected by disruptive and consolidating AI. 
Industry-level AI exposure was measured by calculating the proportion of tasks within each industry that could be affected by either disruptive or consolidating AI technologies (see \nameref{sec:methods}).

A positive correlation was observed between vacancy rates and AI exposure, with a stronger association in the case of disruptive AI. After removing one outlier, accommodation and food services, which deviated by more than two standard deviations from the regression line, the correlation coefficient for disruptive AI reached $r = 0.73$ $(p < 0.05)$. In contrast, the correlation for consolidating AI was more modest and not statistically significant $r = 0.36$ $(p > 0.05$).

\begin{figure}[t]
    \centering
    \includegraphics[width=0.9\linewidth]{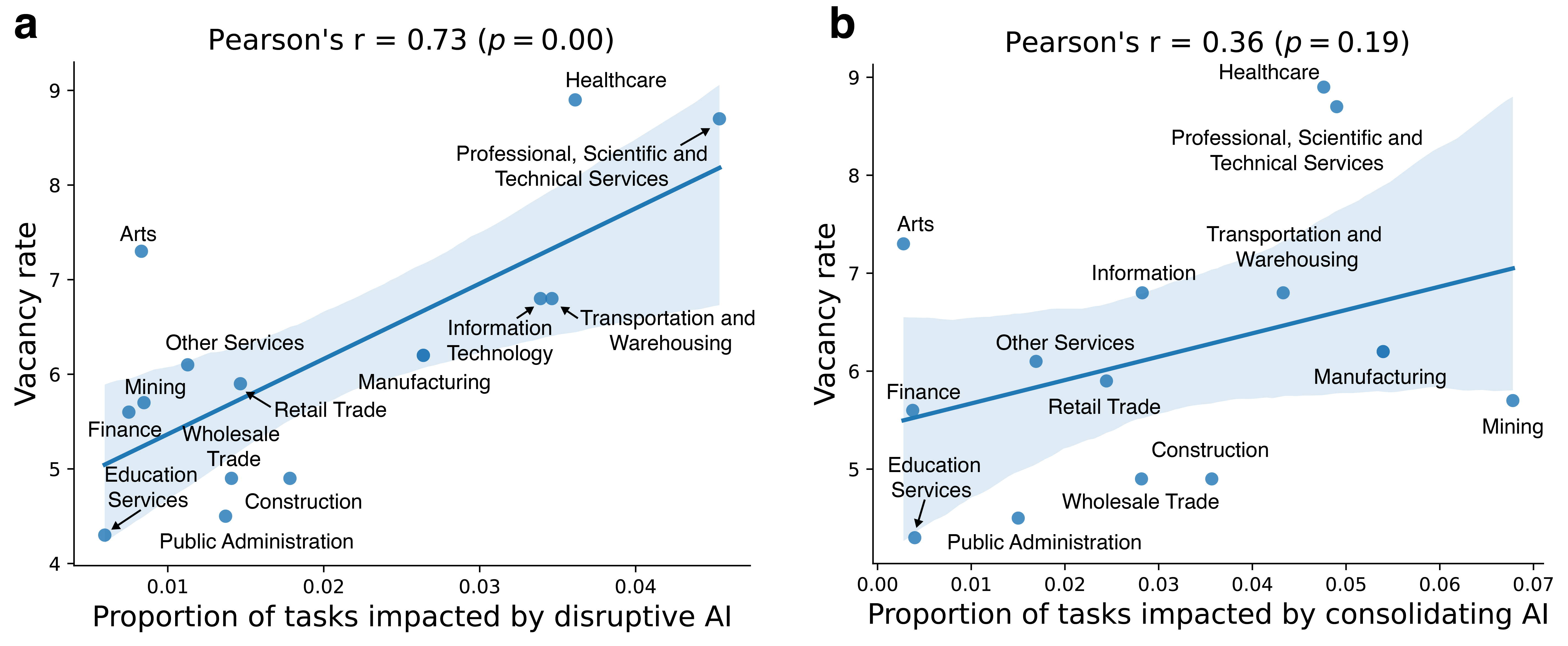}
    \caption{ \textbf{Industry-level relationship between vacancy rates and AI impact of disruptive \emph{vs.} consolidating AI.} A stronger positive correlation is observed between vacancy rates and impact of disruptive AI compared to consolidating AI. This suggests that disruptive AI patents may serve as a more meaningful indicator of labor market shortages. The accommodation and food services sector, which fell more than two standard deviations from the fitted regression line, was treated as an outlier and excluded from this analysis. A version of the figure including this sector is available in the Supplementary Information.}
    \label{fig:vacancy_rate}
\end{figure}
\section*{Discussion}

Our findings show that AI does not affect all job tasks in the same way. As opposed to previous work, by separating AI innovations into two types, disruptive and consolidating, we identified important differences in how they relate to work. Disruptive AI is more likely to affect mental and unpredictable tasks. These tasks are often found in fast-moving, service-based sectors such as information technology, education, and scientific services. However, disruptive AI does not yet appear to affect collaborative work, suggesting that teamwork and social interaction remain difficult for current AI systems to replicate. In contrast, consolidating AI affects physical, predictable, and individual tasks. It is most common in industries that already rely on automation such as manufacturing, construction, and agriculture. This supports earlier research showing that automation has long been part of these sectors and that new AI tools continue this trend by improving existing workflows~\cite{webb2019impact}.

Labor shortages also seem linked to the rise of disruptive AI in the job market. Both disruptive and consolidating AI affect sectors that already face worker gaps, but this link is stronger for disruptive AI. Since automation often grows when there are not enough workers~\cite{graetz2018robots,acemoglu2020robots}, disruptive advances may be more common in these sectors, where new tools help fill the gaps left by too few people.

We also found a strong geographic divide. Disruptive AI patents are concentrated on the U.S. coasts, especially in states like California, New York, and Virginia. These areas have strong innovation networks and are already known for their leadership in technology~\cite{bonaventura2021predicting, acs2002patents, mikhaylov2019coastal}.  Meanwhile, consolidating AI patents are more common in the Midwest and central states, where industries are more focused on physical production.

Our work adds to earlier studies that describe how AI affects different types of work. While past research has shown that unpredictable, cognitive, and collaborative tasks are less likely to be automated~\cite{autor2013tasks,autor2003skill,wilson2018collaborative,davenport2016only,mckinsey2023vitalskills}, our results refine that view. We show that disruptive AI is starting to reach complex cognitive tasks, but not yet collaborative ones. This suggests that social and team-based work may remain out of reach for AI, at least for now.

We also provide new, large-scale evidence to support the idea that disruptive and consolidating technologies play different roles in shaping the economy. This echoes long-standing theories in innovation research, especially Schumpeter's concept of ``creative destruction''~\cite{schumpeter1942capitalism}, highlighting how innovations can either destabilize or enhance existing industries. 
New technologies change the way work is done, often replacing old ways of doing things. Our task-level data shows how this dynamic plays out across occupations, sectors and regions.

\subsection*{Implications}

Our task classification framework can help businesses and policymakers respond to the changing role of AI in the workplace. For businesses, this approach can support task audits that clarify where AI is likely to help and where human work remains essential. Companies in high-disruption sectors can prepare by training workers in decision-making and other cognitive skills, while companies in more traditional sectors may need to focus on reskilling workers for new roles created by automation.

For policymakers, our findings suggest that AI strategies should be tailored by industry and geography. In coastal states where disruptive AI is growing, policies that support innovation and lifelong learning will be critical. In central states where consolidating AI dominates, there is a need for reskilling programs that help workers adapt to more automated environments.

\subsection*{Limitations and Future Work}

Our study has three main limitations.  First, our patent data covers 2015 to 2019. As AI evolves rapidly, newer patents may reflect different patterns. Ongoing analysis is needed to keep up with current trends.

Second, we used both GPT-based models to label the characteristics of the task and human annotators to validate these labels. While agreement was high, both methods have known biases~\cite{ferrara2023should}. Future work should examine discrepancies and test labels against actual labor outcomes.

Finally, we did not directly test whether tasks linked to AI actually led to job loss or change. Linking our patent task framework to real-world labor outcomes, such as wage shifts or unemployment~\cite{frank2023aiexposure}, would be a key next step.

\section*{Methods}
\label{sec:methods}
\subsection*{Datasets}

\noindent\textbf{Occupational Tasks.}
We gathered detailed task descriptions for occupations from version 26.3 of the O*NET database~\cite{onet_archive} released in May 2022, which contains 17,879 unique tasks. 
\smallskip

\noindent \textbf{Patents.} We initially retrieved 6,740,307 patents granted by the United States Patent and Trademark Office (USPTO) up to 2022 from PatentsView~\cite{uspto_patentsview}. After calculating the disruption index quantifying the disruptive characteristics of patents, which is described in the following section, using the entire citation dataset, we focused our analysis on patents published from 2015 onward; a year marked by key AI advancements, including breakthroughs in deep learning architectures such as residual networks and generative adversarial networks, which began reshaping industries, particularly in speech recognition and deep learning~\cite{clark20152015}. We also excluded patents published between 2020 and 2022 from the analysis because the disruption index relies on citation data, which takes time to accumulate after the patent is published. Patents granted more recently have had not enough time to be cited by other patents, which could lead to incomplete results. Then, to reduce the influence of outliers, we excluded patents with fewer than three citations and those without references. Finally, we narrowed our selection to AI-related patents by filtering for AI-specific keywords or CPC codes from the PATENTSCOPE AI Index, following the methodology outlined by Septiandri et al.~\cite{wipo2019wipo,septiandri2023impact} The keywords are: machine learning, deep learning, artificial intelligence, reinforcement learning, neural network, image recognition, computer vision, natural language processing, computational linguistics, speech processing, control methods, knowledge representation, planning, predictive analytics, and robot.

\begin{figure}[t!]
    \centering
    \includegraphics[width=0.8\linewidth]{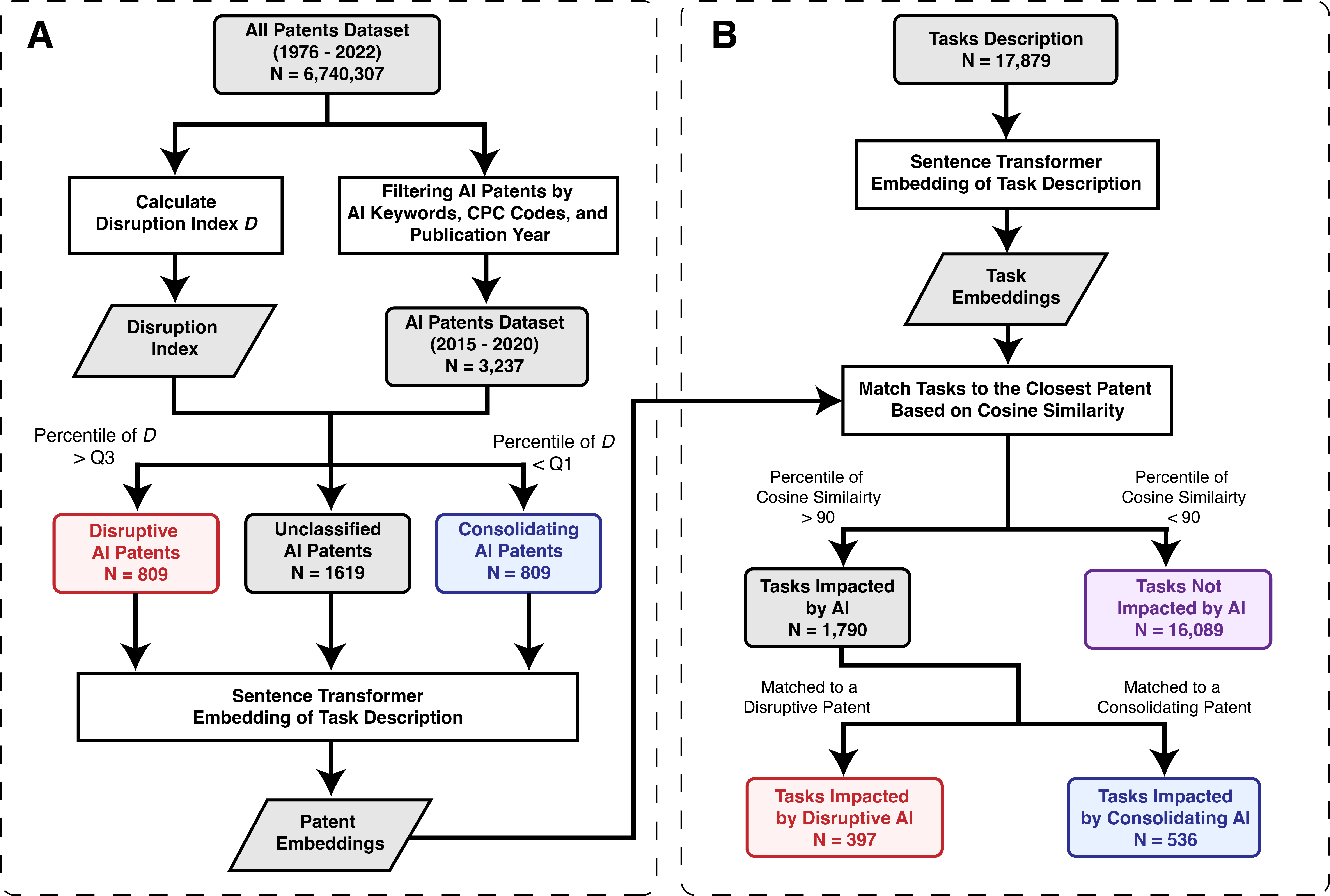}
    \caption{ \textbf{Overview of the Classification Process for Patents and Their Impacted Tasks.} (A) AI-related patents were classified based on their Funk and Owen-Smith's disruption index~\cite{funk2017dynamic}. (B) Task descriptions were matched to patent abstracts based on the highest cosine similarity of their embedding vectors. Note that a single patent can be matched to multiple tasks. Then tasks were classified as `not impacted by AI' if the highest cosine similarity was below the 90\textsuperscript{th} percentile. Otherwise, tasks are classified as `impacted by AI'. Tasks impacted by AI are again classified as `impacted by disruptive AI,' if they are matched to a disruptive AI patent, while those matched to a consolidating AI patent were classified as `impacted by consolidating AI.'}
    \label{fig:schematics}
\end{figure}

\subsection{Classifying Tasks Based on Impact from Disruptive AI and Consolidating AI}
\label{sec:method_class_task}

We classified tasks based on the estimated impact from AI technologies. To do so, as we explain below, we first classified the AI patents as disruptive and consolidating, and then we matched  these patents to occupational tasks by comparing the cosine similarity of their embedding vectors. After matching the patents to occupational tasks having the highest cosine similarity, we classified tasks in either impacted by AI or not impacted by AI depending on their highest similarity scores. Finally, tasks impacted by AI are further classified to tasks impacted by disruptive AI and tasks impacted by consolidating AI depending on the patents matched to the tasks. The overall process is described in Figure~\ref{fig:schematics}.
\smallskip

\subsubsection{Classification of Disruptive and Consolidating AI Patents}
\label{sec:method_class_AI}

We classified each AI patent as either disruptive or consolidating using the disruption index developed by Funk and Owen-Smith~\cite{funk2017dynamic,wu2019large,park2023papers}. This index measures how much a patent changes the direction of future inventions by tracking citation patterns.

The disruption index for a patent $f$ is calculated as:
\begin{equation}
    D_f = \frac{n_i - n_j}{n_i + n_j + n_k},
\end{equation}
where $n_i$ is the number of later patents that cite only the focal patent $f$, $n_j$ is the number that cite both $f$ and its references, and $n_k$ is the number that cite only the references. A higher value means the patent shifts future work away from prior technologies (disruptive), while a lower value means it builds on and strengthens existing ideas (consolidating).

We labeled patents as disruptive, if their index was in the top 25\%; and as consolidating, if it was in the bottom 25\%. We chose these thresholds to focus on clear examples of each type. To check the reliability of this method, we reviewed two things: (1) whether patents identified in prior research as disruptive or consolidating appeared in the expected quartiles~\cite{funk2017dynamic}, and (2) whether a small sample of top-cited patents near each cutoff fit the expected pattern. Specifically, we reviewed the five most-cited patents between the 20th–25th percentiles and the five between the 75th–80th percentiles. These examples supported our classification (see Tables ~\ref{tab:reppat}--\ref{tab:patent_disruption_75_80} in the Supplementary Information). While some patents in the middle of the distribution may be harder to classify, this method gives us a clear distinction between disruptive and consolidating innovations for our analysis.

\smallskip

\subsubsection{Matching tasks to AI patents}
\label{sec:matching}
To determine the extent to which tasks are impacted by AI technology, we employed a sentence transformer model to compare the similarity between task descriptions and AI patent abstract, based on the methodology of Septiandri et al.~\cite{septiandri2023impact}. Specifically, we selected the Sentence-T5-XL model, which has been shown to excel at a range of language tasks, including classification and similarity comparisons~\cite{ni2021sentence}. For each task, we selected the patent with the highest similarity score to estimate the AI potential impact on that task, reflecting its alignment with AI-driven innovations. Following Septiandri et al.~\cite{septiandri2023impact}, tasks with a cosine similarity above the 90\textsuperscript{th} percentile were classified as impacted by AI, while those below this threshold were categorized as not impacted by AI.

We then further refined our classification by examining the type of patents that had the highest cosine similarity with the given AI-impacted tasks. If the patent was identified as disruptive, the task was classified as impacted by disruptive AI. Conversely, if the top-matching patent was classified as consolidating, the task was categorized as impacted by consolidating AI. These processes result in 397 tasks impacted by disruptive patents, 536 tasks impacted by consolidating patents, and 17,548 tasks not impacted by AI.

\subsection{Classifying Tasks Based on Common Job Characteristics}
\label{subsec:classify_tasks_rubric}

\subsubsection{Identifying Key Dimensions of Job Tasks from the Literature}

Many studies in labor economics and sociology have shown that certain types of tasks are more likely to be affected by new technologies, including AI~\cite{autor2003skill, autor2015untangling, autor2013growth, webb2019impact, josten2022automation, frey2017future, felten2021occupational}. 

To identify which task features matter most, we conducted a scoping review using Web of Science. We searched article titles using targeted keywords, limited the results to English-language publications in relevant fields, and reviewed 181 papers. From these, we selected 23 studies that focused on job content, specifically, task characteristics rather than broad categories like high-skilled or low-skilled workers. These papers examined how technology affects jobs, with a focus on skill demands, job polarization, wage shifts, and gender differences~\cite{goos2014explaining, combemale2021not, kikuchi2024automation, UPRETI2022dynamics, EganadelSol2022automation, vannutelli2022routine-biased}.

From this literature, we identified three core dimensions that influence a task's exposure to AI: predictability, mental \emph{vs.} physical effort, and individual \emph{vs.} collaborative work.

\noindent
\textit{Predictability.} Many papers use the routine \emph{vs.} non-routine distinction to describe how technology affects jobs. Tasks with predictable, rule-based procedures are easier to automate~\cite{autor2003skill, autor2013growth, Hoskins2002aggregation}. Studies show that computerization tends to replace routine work but complements unpredictable tasks that require judgment or social interaction~\cite{black2010explaining, combemale2021not}. Even some non-routine tasks can be automated if they are simple or structured~\cite{colin2017complex, frey2017future}.

\noindent
\textit{Mental vs. physical.} Researchers also separate tasks based on whether they require thinking or physical actions. Cognitive tasks often involve reasoning, analysis, or complex problem-solving. These tasks are harder to automate, especially when they depend on judgment or experience~\cite{gray2013taking, tschang2021artificial, webb2019impact}. Physical tasks, in contrast, are more often automated through robotics, especially in manufacturing~\cite{josten2022automation}.

\noindent
\textit{Individual vs. collaborative.} Tasks that rely on teamwork or social interaction—such as negotiation, empathy, or coordination—remain difficult for AI~\cite{deming2017growing, evans2024accelerated}. Studies also highlight the value of soft skills like communication and collaboration, which protect jobs from automation~\cite{colombo2019aimeets}. However, some recent work suggests that even complex and interactive jobs may face automation risks as technologies improve~\cite{montobbio2024labour}.

Based on these findings, we used three binary dimensions to classify job tasks: (1) predictable \emph{vs.} unpredictable, (2) mental \emph{vs.} physical, and (3) individual \emph{vs.} collaborative.

\subsubsection{Collection of GPT-4o Annotations for Type of Tasks Using the Proposed Rubric} 
\label{sec:gpt-4osample}
We asked both GPT-4o to classify each task according to the characteristics outlined in the rubric developed in the previous section. Given the large number of tasks, especially because of tasks not impacted by AI which is over 16,000, conducting a survey for every task was impractical. Therefore, we selected a representative subset of tasks not impacted by AI for annotation consisting of 933 tasks. This number matches the combined total of tasks impacted by both disruptive and consolidating AI.

To select representative tasks, we first built networks of tasks not impacted by AI. In each network, nodes represented tasks, and edges were weighted by the cosine similarity of their sentence embeddings, generated using the Sentence Transformer model. Only edges in the top 95\textsuperscript{th} percentile of similarity were retained to ensure meaningful connections between tasks. We then applied the Louvain algorithm to identify clusters of similar tasks within each network~\cite{blondel2008fast} (see \nameref{sub:network_louvain} in the Supplementary Information for the details of network construction.)

For GPT-4o annotations, we crafted a comprehensive prompt that included a task description, definitions for each classification element in the rubric (e.g., predictable, unpredictable), and annotated examples for guidance. 

The structured prompt ensured that GPT-4 applied the rubric consistently across different representative tasks. The full prompt used for GPT-4 is provided in the  Supplementary Information under \nameref{sec:full_task_prompts}.
\smallskip

\subsubsection*{Validation of GPT-4o Annotations Using Human Annotation} We evaluated the GPT-4 annotations by comparing them to human annotations. Given the large number of tasks involved in the GPT-4 annotations (1,866 tasks in total), it was impractical to conduct a survey for every task. Therefore, following the previous method of sampling representative sets of tasks not impacted by AI, we selected a smaller subset of tasks for further analysis. Specifically, we chose approximately 5 \% of the tasks impacted by disruptive AI and consolidating AI, resulting in 20 tasks impacted by disruptive AI and 28 tasks impacted by consolidating AI. For tasks not impacted by AI, we selected a similar number, totaling 29 tasks. 

For human annotations, we developed and deployed a survey where participants were provided with the industry and occupation of the tasks to aid in their understanding but were not given example annotations to reduce potential bias. 

Participants were recruited via Prolific, and each task was annotated by three different crowd workers to ensure reliability~\cite{prolific}. After collecting the responses, the majority answer was selected as the final annotation for each task, ensuring that the most commonly agreed-upon classification was used in our analysis.

After collecting both human and GPT-4o annotations and calculating the alignment rate between these two annotation sets, we further identified tasks where there was misalignment between the two sets of annotations and validated which annotation is more reliable. Three authors independently annotated the tasks with misaligned annotations. If a misalignment persisted after this step, the authors discussed the differing annotations and reached a consensus to determine the final classification. This process ensured that the final annotations were accurate and reflected the collective judgment of the research team.
\smallskip

\subsubsection*{Significance Testing of Task Characteristics} To assess the significance of the differences observed in task characteristics after human annotation and survey, we implemented a null model to evaluate whether the observed ratios deviated significantly from what would be expected by chance. We constructed a null model by randomly shuffling the task characteristics across the dataset. This process involved reassigning the characteristics to tasks randomly, simulating a scenario where any observed patterns could be due to chance. This randomization was performed 500 times to create a distribution of ratios under the null hypothesis. For each iteration of the null model, we calculated the proportion of tasks with specific characteristics (e.g., automation potential) and compiled these proportions into a null distribution.

To assess how the observed ratios deviated from this null distribution, we computed the $z$-score for each observed characteristic $c$ ratio as follows: 

\begin{equation}
    z_{c} = \frac{\text{Observed Ratio of $c$} - \text{Mean of Null Ratios}}{\text{Standard Deviation of Null Ratios}} 
\end{equation}

\subsection*{Industry Sectoral and Geographical Level Analysis of Patents Impacting Occupational Tasks}

\subsubsection{Industry Level}
\label{sub:industry}

To assess how the impact of disruptive and consolidating AI patents differs from that of general AI across various industries, we began by calculating the proportions of AI patents that affect occupational tasks within each category (disruptive and consolidating). We then compared these proportions to the overall industry distribution of tasks impacted by general AI by subtracting them from the latter. This calculation enabled us to identify the deviation, or the degree to which the influence of disruptive and consolidating AI patents in specific industries diverges from the broader trends in AI patent distribution. Specifically, this deviation is defined as:
\begin{align}
    \Delta P_D^I(X) &= P_D^I(X) - P^I(X) \\
    \Delta P_C^I(X) &= P_C^I(X) - P^I(X), 
\end{align}

where $P_D^I(X)$ and $P_C^I(X)$ indicates the proportion of the industry sector $X$ in tasks impacted by disruptive AI patents and consolidating AI patents, and $P^I(X)$ indicates the proportion of the industry sector $X$ in tasks impacted by all AI patents. Formally, these proportions are defined as:

\begin{align}
    P_D^I &= \frac{T_D^I}{T_D} \\
    P_C^I &= \frac{T_C^I}{T_C} \\
    P^I   &= \frac{T^I}{T}.
\end{align}

\noindent
Here, $T_D^I$ is the number of the tasks impacted by disruptive AI in industry $I$ and $T_D$ is the total number of tasks impacted by disruptive AI across all industries. Similarly, $T_C^I$ represents the number of tasks impacted by consolidating AI in industry $I$, $T_C$ the total across all industries. $T^I$ is the number of AI-impacted tasks in industry $I$, and $T$ is the total number of AI-impacted tasks across all industries.

\smallskip

\subsubsection*{State Level}
For the state-level analysis of patent impact on occupational tasks, we calculated the difference in the proportion of AI patents published in each state between disruptive and consolidating AI patents that impact occupational tasks as

\begin{equation}
    \Delta P^S(X) = P^S_D(X) - P^S_C(X).
\end{equation}

In this calculation, we again allowed for duplication when a single patent is matched to multiple tasks. Here, $P^S_D(X)$ indicates the proportion of patents published in state $X$ among disruptive AI patents having an impact on occupational tasks and $P^S_C(X)$ is the proportion of patents published in state $X$ among consolidating AI patents.

\subsection*{Quantifying Sectoral AI Exposure}

The extent to which each industry is affected by disruptive or consolidating AI technologies was quantified as the proportion of tasks within that industry impacted by either type of AI. Formally, for each industry $i$,  this is defined as:

\begin{equation}
\frac{T_i^{\text{Disruptive}}}{T_i}
\qquad \text{and} \qquad
\frac{T_i^{\text{Consolidating}}}{T_i}
\end{equation}

\revision{where $T_i$ is the total number of tasks associated with industry $i$,  $T_i^{\text{Disruptive}}$  is the number of tasks in industry $i$ impacted by disruptive AI, and $T_i^{\text{Consolidating}}$ is the number of tasks in industry $i$ impacted by consolidating AI. }

\section*{Acknowledgements}
We also thank to Morgan Frank and
Edyta Paulina Bogucka for their valuable comments.

\section*{Funding}
Nokia Bell Labs supported this work through author salaries, but had no involvement in the study design, data collection and analysis, decision to publish, or manuscript preparation.

\section*{Author contributions statement}
MK conceptualized the ideas, collected the data, conducted the analysis, and wrote the first draft of the manuscript. MC and SS conceptualized the ideas, designed and executed the survey, conceived the experiments, and provided feedback on the manuscript. DQ conceptualized the ideas, conceived the experiments, and provided feedback on the manuscript. YYA provided feedback on the manuscript.

\section*{Data and Code availability}
Data and code necessary to reproduce the analyses in this study is available in the project's page at 
\url{https://anonymous.4open.science/r/Impact-of-Disruptive-AI-on-Task-D67C/README.md}

\bibliography{main}

\begin{thebibliography}{10}
\urlstyle{rm}
\expandafter\ifx\csname url\endcsname\relax
  \def\url#1{\texttt{#1}}\fi
\expandafter\ifx\csname urlprefix\endcsname\relax\def\urlprefix{URL }\fi
\expandafter\ifx\csname doiprefix\endcsname\relax\def\doiprefix{DOI: }\fi
\providecommand{\bibinfo}[2]{#2}
\providecommand{\eprint}[2][]{\url{#2}}

\bibitem{demirci2023ai}
\bibinfo{author}{Demirci, O.}, \bibinfo{author}{Hannane, J.} \& \bibinfo{author}{Zhu, X.}
\newblock \bibinfo{title}{Who is ai replacing? the impact of generative ai on online freelancing platforms}.
\newblock \bibinfo{type}{Working Paper} \bibinfo{number}{11276}, \bibinfo{institution}{CESifo} (\bibinfo{year}{2023}).
\newblock \bibinfo{note}{CESifo Working Paper No. 11276}.

\bibitem{liu2023generate}
\bibinfo{author}{Liu, J.}, \bibinfo{author}{Xu, X.}, \bibinfo{author}{Li, Y.} \& \bibinfo{author}{Tan, Y.}
\newblock \bibinfo{journal}{\bibinfo{title}{``generate" the future of work through ai: Empirical evidence from online labor markets}}.
\newblock {\emph{\JournalTitle{arXiv preprint arXiv:2308.05201}}}  (\bibinfo{year}{2023}).

\bibitem{noy2023experimental}
\bibinfo{author}{Noy, S.} \& \bibinfo{author}{Zhang, W.}
\newblock \bibinfo{journal}{\bibinfo{title}{Experimental evidence on the productivity effects of generative artificial intelligence}}.
\newblock {\emph{\JournalTitle{Science}}} \textbf{\bibinfo{volume}{381}}, \bibinfo{pages}{187--192}, \doiprefix\url{10.1126/science.adh2586} (\bibinfo{year}{2023}).

\bibitem{agrawal2022chatgpt}
\bibinfo{author}{Agrawal, A.}, \bibinfo{author}{Gans, J.} \& \bibinfo{author}{Goldfarb, A.}
\newblock \bibinfo{journal}{\bibinfo{title}{{ChatGPT and How AI Disrupts Industries}}}.
\newblock {\emph{\JournalTitle{Harvard Business Review}}}  (\bibinfo{year}{2022}).

\bibitem{gmyrek2023genai}
\bibinfo{author}{Gmyrek, P.}, \bibinfo{author}{Berg, J.} \& \bibinfo{author}{Bescond, D.}
\newblock \emph{\bibinfo{title}{Generative AI and jobs: a global analysis of potential effects on job quantity and quality}} (\bibinfo{publisher}{ILO}, \bibinfo{year}{2023}).

\bibitem{cappelli2024will}
\bibinfo{author}{Cappelli, P.}, \bibinfo{author}{Tambe, P.~S.} \& \bibinfo{author}{Yakubovich, V.}
\newblock \bibinfo{journal}{\bibinfo{title}{Will large language models really change how work is done?}}
\newblock {\emph{\JournalTitle{MIT Sloan Management Review}}} \bibinfo{pages}{19--24} (\bibinfo{year}{2024}).
\newblock \bibinfo{note}{Included in the Companion Article Pack}.

\bibitem{morandini2023impact}
\bibinfo{author}{Morandini, S.} \emph{et~al.}
\newblock \bibinfo{journal}{\bibinfo{title}{The impact of artificial intelligence on workers’ skills: Upskilling and reskilling in organisations}}.
\newblock {\emph{\JournalTitle{Informing Science}}} \textbf{\bibinfo{volume}{26}}, \bibinfo{pages}{39--68}, \doiprefix\url{10.28945/5078} (\bibinfo{year}{2023}).

\bibitem{tushman1986technological}
\bibinfo{author}{Tushman, M.~L.} \& \bibinfo{author}{Anderson, P.}
\newblock \bibinfo{journal}{\bibinfo{title}{Technological discontinuities and organizational environments}}.
\newblock {\emph{\JournalTitle{Administrative Science Quarterly}}} \textbf{\bibinfo{volume}{31}}, \bibinfo{pages}{439--465} (\bibinfo{year}{1986}).

\bibitem{baker2012technology}
\bibinfo{author}{Baker, J.}
\newblock \bibinfo{title}{The technology--organization--environment framework}.
\newblock In \bibinfo{editor}{Dwivedi, Y.~K.}, \bibinfo{editor}{Wade, M.~R.} \& \bibinfo{editor}{Schneberger, S.~L.} (eds.) \emph{\bibinfo{booktitle}{Information Systems Theory: Explaining and Predicting Our Digital Society, Vol. 1}}, \bibinfo{pages}{231--245}, \doiprefix\url{10.1007/978-1-4419-6108-2_12} (\bibinfo{publisher}{Springer}, \bibinfo{address}{New York, NY}, \bibinfo{year}{2012}).

\bibitem{hci_disruption}
\bibinfo{author}{Chen, Z.} \& \bibinfo{author}{Li, Y.}
\newblock \bibinfo{title}{The sharply decreasing disruptiveness of hci}.
\newblock In \emph{\bibinfo{booktitle}{Proceedings of the 2025 CHI Conference on Human Factors in Computing Systems}}, CHI '25, \doiprefix\url{10.1145/3706598.3713917} (\bibinfo{publisher}{Association for Computing Machinery}, \bibinfo{address}{New York, NY, USA}, \bibinfo{year}{2025}).

\bibitem{funk2017dynamic}
\bibinfo{author}{Funk, R.~J.} \& \bibinfo{author}{Owen-Smith, J.}
\newblock \bibinfo{journal}{\bibinfo{title}{A dynamic network measure of technological change}}.
\newblock {\emph{\JournalTitle{Management science}}} \textbf{\bibinfo{volume}{63}}, \bibinfo{pages}{791--817}, \doiprefix\url{10.1287/mnsc.2015.2366} (\bibinfo{year}{2017}).

\bibitem{wu2019large}
\bibinfo{author}{Wu, L.}, \bibinfo{author}{Wang, D.} \& \bibinfo{author}{Evans, J.~A.}
\newblock \bibinfo{journal}{\bibinfo{title}{Large teams develop and small teams disrupt science and technology}}.
\newblock {\emph{\JournalTitle{Nature}}} \textbf{\bibinfo{volume}{566}}, \bibinfo{pages}{378--382}, \doiprefix\url{10.1038/s41586-019-0941-9} (\bibinfo{year}{2019}).

\bibitem{jesuthasan2021work}
\bibinfo{author}{Jesuthasan, R.} \& \bibinfo{author}{Boudreau, J.}
\newblock \bibinfo{journal}{\bibinfo{title}{Work without jobs}}.
\newblock {\emph{\JournalTitle{MIT Sloan Management Review}}} \textbf{\bibinfo{volume}{62}}, \bibinfo{pages}{1--5} (\bibinfo{year}{2021}).

\bibitem{autor2013growth}
\bibinfo{author}{Autor, D.~H.} \& \bibinfo{author}{Dorn, D.}
\newblock \bibinfo{journal}{\bibinfo{title}{The growth of low-skill service jobs and the polarization of the us labor market}}.
\newblock {\emph{\JournalTitle{American economic review}}} \textbf{\bibinfo{volume}{103}}, \bibinfo{pages}{1553--1597} (\bibinfo{year}{2013}).

\bibitem{autor2015untangling}
\bibinfo{author}{Autor, D.~H.}, \bibinfo{author}{Dorn, D.} \& \bibinfo{author}{Hanson, G.~H.}
\newblock \bibinfo{journal}{\bibinfo{title}{Untangling trade and technology: Evidence from local labour markets}}.
\newblock {\emph{\JournalTitle{The Economic Journal}}} \textbf{\bibinfo{volume}{125}}, \bibinfo{pages}{621--646}, \doiprefix\url{10.1111/ecoj.12245} (\bibinfo{year}{2015}).
\newblock \eprint{https://academic.oup.com/ej/article-pdf/125/584/621/26438955/ej0621.pdf}.

\bibitem{deming2017growing}
\bibinfo{author}{Deming, D.~J.}
\newblock \bibinfo{journal}{\bibinfo{title}{The growing importance of social skills in the labor market*}}.
\newblock {\emph{\JournalTitle{The Quarterly Journal of Economics}}} \textbf{\bibinfo{volume}{132}}, \bibinfo{pages}{1593--1640}, \doiprefix\url{10.1093/qje/qjx022} (\bibinfo{year}{2017}).
\newblock \eprint{https://academic.oup.com/qje/article-pdf/132/4/1593/30637898/qjx022.pdf}.

\bibitem{septiandri2023impact}
\bibinfo{author}{Septiandri, A.~A.}, \bibinfo{author}{Constantinides, M.} \& \bibinfo{author}{Quercia, D.}
\newblock \bibinfo{journal}{\bibinfo{title}{The impact of ai innovations on us occupations}}.
\newblock {\emph{\JournalTitle{arXiv preprint arXiv:2312.04714}}}  (\bibinfo{year}{2023}).

\bibitem{webb2019impact}
\bibinfo{author}{Webb, M.}
\newblock \bibinfo{journal}{\bibinfo{title}{The impact of artificial intelligence on the labor market}}.
\newblock {\emph{\JournalTitle{Available at SSRN 3482150}}}  (\bibinfo{year}{2019}).

\bibitem{autor2024applying}
\bibinfo{author}{Autor, D.}
\newblock \bibinfo{title}{Applying ai to rebuild middle class jobs}.
\newblock \bibinfo{type}{Tech. Rep.}, \bibinfo{institution}{National Bureau of Economic Research} (\bibinfo{year}{2024}).

\bibitem{Constantinides2025AI}
\bibinfo{author}{Constantinides, M.} \& \bibinfo{author}{Quercia, D.}
\newblock \bibinfo{title}{{AI, Jobs, and the Automation Trap: Where Is HCI?}}
\newblock In \emph{\bibinfo{booktitle}{Proceedings of the 4th ACM Annual Symposium on Human-Computer Interaction for Work (CHIWORK '25)}}, \bibinfo{pages}{8} (\bibinfo{year}{2025}).

\bibitem{graetz2018robots}
\bibinfo{author}{Graetz, G.} \& \bibinfo{author}{Michaels, G.}
\newblock \bibinfo{journal}{\bibinfo{title}{Robots at work}}.
\newblock {\emph{\JournalTitle{Review of economics and statistics}}} \textbf{\bibinfo{volume}{100}}, \bibinfo{pages}{753--768} (\bibinfo{year}{2018}).

\bibitem{acemoglu2020robots}
\bibinfo{author}{Acemoglu, D.} \& \bibinfo{author}{Restrepo, P.}
\newblock \bibinfo{journal}{\bibinfo{title}{Robots and jobs: Evidence from us labor markets}}.
\newblock {\emph{\JournalTitle{Journal of political economy}}} \textbf{\bibinfo{volume}{128}}, \bibinfo{pages}{2188--2244} (\bibinfo{year}{2020}).

\bibitem{bonaventura2021predicting}
\bibinfo{author}{Bonaventura, M.}, \bibinfo{author}{Aiello, L.~M.}, \bibinfo{author}{Quercia, D.} \& \bibinfo{author}{Latora, V.}
\newblock \bibinfo{journal}{\bibinfo{title}{Predicting urban innovation from the us workforce mobility network}}.
\newblock {\emph{\JournalTitle{Humanities and Social Sciences Communications}}} \textbf{\bibinfo{volume}{8}}, \bibinfo{pages}{1--9} (\bibinfo{year}{2021}).

\bibitem{acs2002patents}
\bibinfo{author}{Acs, Z.~J.}, \bibinfo{author}{Anselin, L.} \& \bibinfo{author}{Varga, A.}
\newblock \bibinfo{journal}{\bibinfo{title}{Patents and innovation counts as measures of regional production of new knowledge}}.
\newblock {\emph{\JournalTitle{Research policy}}} \textbf{\bibinfo{volume}{31}}, \bibinfo{pages}{1069--1085}, \doiprefix\url{10.1016/S0048-7333(01)00184-6} (\bibinfo{year}{2002}).

\bibitem{mikhaylov2019coastal}
\bibinfo{author}{Mikhaylov, A.~S.}
\newblock \bibinfo{journal}{\bibinfo{title}{Coastal agglomerations and the transformation of national innovation spaces}}.
\newblock {\emph{\JournalTitle{Baltic Region}}} \textbf{\bibinfo{volume}{11}}, \bibinfo{pages}{29--42}, \doiprefix\url{10.5922/2079-8555-2019-1-3} (\bibinfo{year}{2019}).

\bibitem{autor2013tasks}
\bibinfo{author}{Autor, D.~H.} \& \bibinfo{author}{Handel, M.~J.}
\newblock \bibinfo{journal}{\bibinfo{title}{Putting tasks to the test: Human capital, job tasks, and wages}}.
\newblock {\emph{\JournalTitle{Journal of Labor Economics}}} \textbf{\bibinfo{volume}{31}}, \bibinfo{pages}{S59--S96}, \doiprefix\url{10.1086/669332} (\bibinfo{year}{2013}).

\bibitem{autor2003skill}
\bibinfo{author}{Autor, D.~H.}, \bibinfo{author}{Levy, F.} \& \bibinfo{author}{Murnane, R.~J.}
\newblock \bibinfo{journal}{\bibinfo{title}{The skill content of recent technological change: An empirical exploration*}}.
\newblock {\emph{\JournalTitle{The Quarterly Journal of Economics}}} \textbf{\bibinfo{volume}{118}}, \bibinfo{pages}{1279--1333}, \doiprefix\url{10.1162/003355303322552801} (\bibinfo{year}{2003}).
\newblock \eprint{https://academic.oup.com/qje/article-pdf/118/4/1279/5427313/118-4-1279.pdf}.

\bibitem{wilson2018collaborative}
\bibinfo{author}{Wilson, H.~J.} \& \bibinfo{author}{Daugherty, P.~R.}
\newblock \bibinfo{journal}{\bibinfo{title}{Collaborative intelligence: Humans and ai are joining forces}}.
\newblock {\emph{\JournalTitle{Harvard Business Review}}} \textbf{\bibinfo{volume}{96}}, \bibinfo{pages}{114--123} (\bibinfo{year}{2018}).

\bibitem{davenport2016only}
\bibinfo{author}{Davenport, T.~H.} \& \bibinfo{author}{Kirby, J.}
\newblock \emph{\bibinfo{title}{Only humans need apply: Winners and losers in the age of smart machines}} (\bibinfo{publisher}{Harper Business New York}, \bibinfo{year}{2016}).

\bibitem{mckinsey2023vitalskills}
\bibinfo{author}{{McKinsey \& Company}}.
\newblock \bibinfo{journal}{\bibinfo{title}{Building the vital skills for the future of work in operations}}.
\newblock {\emph{\JournalTitle{McKinsey \& Company}}}  (\bibinfo{year}{2023}).
\newblock \bibinfo{note}{Accessed: 2024-08-20}.

\bibitem{schumpeter1942capitalism}
\bibinfo{author}{Schumpeter, J.~A.}
\newblock \bibinfo{journal}{\bibinfo{title}{Capitalism}}.
\newblock {\emph{\JournalTitle{Socialism and Democracy/Harper}}}  (\bibinfo{year}{1942}).

\bibitem{ferrara2023should}
\bibinfo{author}{Ferrara, E.}
\newblock \bibinfo{journal}{\bibinfo{title}{Should chatgpt be biased? challenges and risks of bias in large language models}}.
\newblock {\emph{\JournalTitle{arXiv preprint arXiv:2304.03738}}}  (\bibinfo{year}{2023}).

\bibitem{frank2023aiexposure}
\bibinfo{author}{Frank, M.}, \bibinfo{author}{Ahn, Y.-Y.} \& \bibinfo{author}{Moro, E.}
\newblock \bibinfo{title}{{AI exposure predicts unemployment risk}} (\bibinfo{year}{2023}).
\newblock \eprint{2308.02624}.

\bibitem{onet_archive}
\bibinfo{author}{{National Center for O*NET Development}}.
\newblock \bibinfo{title}{{O*NET® Database Releases Archive}}.
\newblock \bibinfo{howpublished}{\url{www.onetcenter.org/db_releases.html}} (\bibinfo{year}{2024}).
\newblock \bibinfo{note}{Accessed: 2024-08-08}.

\bibitem{uspto_patentsview}
\bibinfo{author}{{U.S. Patent and Trademark Office}}.
\newblock \bibinfo{title}{{Data Download Tables}}.
\newblock \bibinfo{howpublished}{\url{https://patentsview.org/download/data-download-tables}}.
\newblock \bibinfo{note}{Accessed: [date]}.

\bibitem{clark20152015}
\bibinfo{author}{Clark, J.}
\newblock \bibinfo{journal}{\bibinfo{title}{Why 2015 was a breakthrough year in artificial intelligence}}.
\newblock {\emph{\JournalTitle{Bloomberg News}}} \textbf{\bibinfo{volume}{8}} (\bibinfo{year}{2015}).

\bibitem{wipo2019wipo}
\bibinfo{author}{WIPO}.
\newblock \bibinfo{journal}{\bibinfo{title}{Wipo technology trends 2019: Artificial intelligence}}.
\newblock {\emph{\JournalTitle{Geneva: World Intellectual Property Organization}}}  (\bibinfo{year}{2019}).

\bibitem{park2023papers}
\bibinfo{author}{Park, M.}, \bibinfo{author}{Leahey, E.} \& \bibinfo{author}{Funk, R.~J.}
\newblock \bibinfo{journal}{\bibinfo{title}{Papers and patents are becoming less disruptive over time}}.
\newblock {\emph{\JournalTitle{Nature}}} \textbf{\bibinfo{volume}{613}}, \bibinfo{pages}{138--144}, \doiprefix\url{10.1038/s41586-022-05543-x} (\bibinfo{year}{2023}).

\bibitem{ni2021sentence}
\bibinfo{author}{Ni, J.} \emph{et~al.}
\newblock \bibinfo{journal}{\bibinfo{title}{Sentence-t5: Scalable sentence encoders from pre-trained text-to-text models}}.
\newblock {\emph{\JournalTitle{arXiv preprint arXiv:2108.08877}}}  (\bibinfo{year}{2021}).

\bibitem{josten2022automation}
\bibinfo{author}{Josten, C.} \& \bibinfo{author}{Lordan, G.}
\newblock \bibinfo{journal}{\bibinfo{title}{Automation and the changing nature of work}}.
\newblock {\emph{\JournalTitle{Plos one}}} \textbf{\bibinfo{volume}{17}}, \bibinfo{pages}{e0266326} (\bibinfo{year}{2022}).

\bibitem{frey2017future}
\bibinfo{author}{Frey, C.~B.} \& \bibinfo{author}{Osborne, M.~A.}
\newblock \bibinfo{journal}{\bibinfo{title}{The future of employment: How susceptible are jobs to computerisation?}}
\newblock {\emph{\JournalTitle{Technological Forecasting and Social Change}}} \textbf{\bibinfo{volume}{114}}, \bibinfo{pages}{254--280}, \doiprefix\url{https://doi.org/10.1016/j.techfore.2016.08.019} (\bibinfo{year}{2017}).

\bibitem{felten2021occupational}
\bibinfo{author}{Felten, E.}, \bibinfo{author}{Raj, M.} \& \bibinfo{author}{Seamans, R.}
\newblock \bibinfo{journal}{\bibinfo{title}{Occupational, industry, and geographic exposure to artificial intelligence: A novel dataset and its potential uses}}.
\newblock {\emph{\JournalTitle{Strategic Management Journal}}} \textbf{\bibinfo{volume}{42}}, \bibinfo{pages}{2195--2217}, \doiprefix\url{https://doi.org/10.1002/smj.3286} (\bibinfo{year}{2021}).
\newblock \eprint{https://onlinelibrary.wiley.com/doi/pdf/10.1002/smj.3286}.

\bibitem{goos2014explaining}
\bibinfo{author}{Goos, M.}, \bibinfo{author}{Manning, A.} \& \bibinfo{author}{Salomons, A.}
\newblock \bibinfo{journal}{\bibinfo{title}{Explaining job polarization: Routine-biased technological change and offshoring}}.
\newblock {\emph{\JournalTitle{American Economic Review}}} \textbf{\bibinfo{volume}{104}}, \bibinfo{pages}{2509–26}, \doiprefix\url{10.1257/aer.104.8.2509} (\bibinfo{year}{2014}).

\bibitem{combemale2021not}
\bibinfo{author}{Combemale, C.}, \bibinfo{author}{Whitefoot, K.~S.}, \bibinfo{author}{Ales, L.} \& \bibinfo{author}{Fuchs, E. R.~H.}
\newblock \bibinfo{journal}{\bibinfo{title}{Not all technological change is equal: how the separability of tasks mediates the effect of technology change on skill demand}}.
\newblock {\emph{\JournalTitle{Industrial and Corporate Change}}} \textbf{\bibinfo{volume}{30}}, \bibinfo{pages}{1361--1387}, \doiprefix\url{10.1093/icc/dtab026} (\bibinfo{year}{2021}).
\newblock \bibinfo{note}{\_eprint: https://academic.oup.com/icc/article-pdf/30/6/1361/42005347/dtab026.pdf}.

\bibitem{kikuchi2024automation}
\bibinfo{author}{Kikuchi, S.}, \bibinfo{author}{Fujiwara, I.} \& \bibinfo{author}{Shirota, T.}
\newblock \bibinfo{journal}{\bibinfo{title}{Automation and disappearing routine occupations in japan}}.
\newblock {\emph{\JournalTitle{Journal of the Japanese and International Economies}}} \textbf{\bibinfo{volume}{74}}, \bibinfo{pages}{101338}, \doiprefix\url{https://doi.org/10.1016/j.jjie.2024.101338} (\bibinfo{year}{2024}).

\bibitem{UPRETI2022dynamics}
\bibinfo{author}{UPRETI, A.} \& \bibinfo{author}{SRIDHAR, V.}
\newblock \bibinfo{journal}{\bibinfo{title}{The dynamics of task automation and worker adjustment in labor markets: An agent-based approach}}.
\newblock {\emph{\JournalTitle{Advances in Complex Systems}}} \textbf{\bibinfo{volume}{25}}, \bibinfo{pages}{2250005}, \doiprefix\url{10.1142/S0219525922500059} (\bibinfo{year}{2022}).
\newblock \eprint{https://doi.org/10.1142/S0219525922500059}.

\bibitem{EganadelSol2022automation}
\bibinfo{author}{Egana-delSol, P.}, \bibinfo{author}{Bustelo, M.}, \bibinfo{author}{Ripani, L.}, \bibinfo{author}{Soler, N.} \& \bibinfo{author}{Viollaz, M.}
\newblock \bibinfo{journal}{\bibinfo{title}{Automation in latin america: Are women at higher risk of losing their jobs?}}
\newblock {\emph{\JournalTitle{Technological Forecasting and Social Change}}} \textbf{\bibinfo{volume}{175}}, \bibinfo{pages}{121333}, \doiprefix\url{https://doi.org/10.1016/j.techfore.2021.121333} (\bibinfo{year}{2022}).

\bibitem{vannutelli2022routine-biased}
\bibinfo{author}{Vannutelli, S.}, \bibinfo{author}{Scicchitano, S.} \& \bibinfo{author}{Biagetti, M.}
\newblock \bibinfo{journal}{\bibinfo{title}{Routine-biased technological change and wage inequality: do workers’ perceptions matter?}}
\newblock {\emph{\JournalTitle{Eurasian Business Review}}} \textbf{\bibinfo{volume}{12}}, \bibinfo{pages}{409--450}, \doiprefix\url{10.1007/s40821-022-00222-3} (\bibinfo{year}{2022}).

\bibitem{Hoskins2002aggregation}
\bibinfo{author}{Hoskins, M.}
\newblock \bibinfo{journal}{\bibinfo{title}{Aggregation, technological change and the skill structure of the british economy 1951-1991}}.
\newblock {\emph{\JournalTitle{Applied Economics Letters}}} \textbf{\bibinfo{volume}{9}}, \bibinfo{pages}{251--254}, \doiprefix\url{10.1080/13504850110060376} (\bibinfo{year}{2002}).
\newblock \eprint{https://doi.org/10.1080/13504850110060376}.

\bibitem{black2010explaining}
\bibinfo{author}{Black, S.~E.} \& \bibinfo{author}{Spitz-Oener, A.}
\newblock \bibinfo{journal}{\bibinfo{title}{Explaining women's success: Technological change and the skill content of women's work}}.
\newblock {\emph{\JournalTitle{The Review of Economics and Statistics}}} \textbf{\bibinfo{volume}{92}}, \bibinfo{pages}{187--194} (\bibinfo{year}{2010}).

\bibitem{colin2017complex}
\bibinfo{author}{Caines, C.}, \bibinfo{author}{Hoffmann, F.} \& \bibinfo{author}{Kambourov, G.}
\newblock \bibinfo{journal}{\bibinfo{title}{Complex-task biased technological change and the labor market}}.
\newblock {\emph{\JournalTitle{Review of Economic Dynamics}}} \textbf{\bibinfo{volume}{25}}, \bibinfo{pages}{298--319}, \doiprefix\url{https://doi.org/10.1016/j.red.2017.01.008} (\bibinfo{year}{2017}).
\newblock \bibinfo{note}{Special Issue on Human Capital and Inequality}.

\bibitem{gray2013taking}
\bibinfo{author}{Gray, R.}
\newblock \bibinfo{journal}{\bibinfo{title}{Taking technology to task: The skill content of technological change in early twentieth century united states}}.
\newblock {\emph{\JournalTitle{Explorations in Economic History}}} \textbf{\bibinfo{volume}{50}}, \bibinfo{pages}{351--367}, \doiprefix\url{https://doi.org/10.1016/j.eeh.2013.04.002} (\bibinfo{year}{2013}).

\bibitem{tschang2021artificial}
\bibinfo{author}{Tschang, F.~T.} \& \bibinfo{author}{Almirall, E.}
\newblock \bibinfo{journal}{\bibinfo{title}{Artificial intelligence as augmenting automation: Implications for employment}}.
\newblock {\emph{\JournalTitle{Academy of Management Perspectives}}} \textbf{\bibinfo{volume}{35}}, \bibinfo{pages}{642--659}, \doiprefix\url{10.5465/amp.2019.0062} (\bibinfo{year}{2021}).

\bibitem{evans2024accelerated}
\bibinfo{author}{Evans, D.}, \bibinfo{author}{Mason, C.}, \bibinfo{author}{Chen, H.} \& \bibinfo{author}{Reeson, A.}
\newblock \bibinfo{journal}{\bibinfo{title}{Accelerated demand for interpersonal skills in the {Australian} post-pandemic labour market}}.
\newblock {\emph{\JournalTitle{Nature Human Behaviour}}} \textbf{\bibinfo{volume}{8}}, \bibinfo{pages}{32--42}, \doiprefix\url{10.1038/s41562-023-01788-2} (\bibinfo{year}{2024}).

\bibitem{colombo2019aimeets}
\bibinfo{author}{Colombo, E.}, \bibinfo{author}{Mercorio, F.} \& \bibinfo{author}{Mezzanzanica, M.}
\newblock \bibinfo{journal}{\bibinfo{title}{Ai meets labor market: Exploring the link between automation and skills}}.
\newblock {\emph{\JournalTitle{Information Economics and Policy}}} \textbf{\bibinfo{volume}{47}}, \bibinfo{pages}{27--37}, \doiprefix\url{https://doi.org/10.1016/j.infoecopol.2019.05.003} (\bibinfo{year}{2019}).
\newblock \bibinfo{note}{The Economics of Artificial Intelligence and Machine Learning}.

\bibitem{montobbio2024labour}
\bibinfo{author}{Montobbio, F.}, \bibinfo{author}{Staccioli, J.}, \bibinfo{author}{Virgillito, M.~E.} \& \bibinfo{author}{Vivarelli, M.}
\newblock \bibinfo{journal}{\bibinfo{title}{Labour-saving automation: A direct measure of occupational exposure}}.
\newblock {\emph{\JournalTitle{The World Economy}}} \textbf{\bibinfo{volume}{47}}, \bibinfo{pages}{332--361} (\bibinfo{year}{2024}).

\bibitem{blondel2008fast}
\bibinfo{author}{Blondel, V.~D.}, \bibinfo{author}{Guillaume, J.-L.}, \bibinfo{author}{Lambiotte, R.} \& \bibinfo{author}{Lefebvre, E.}
\newblock \bibinfo{journal}{\bibinfo{title}{Fast unfolding of communities in large networks}}.
\newblock {\emph{\JournalTitle{Journal of statistical mechanics: theory and experiment}}} \textbf{\bibinfo{volume}{2008}}, \bibinfo{pages}{P10008}, \doiprefix\url{10.1088/1742-5468/2008/10/P10008} (\bibinfo{year}{2008}).

\bibitem{prolific}
\bibinfo{author}{{Prolific}}.
\newblock \bibinfo{title}{Prolific}.
\newblock \bibinfo{howpublished}{\url{https://www.prolific.co/}} (\bibinfo{year}{2024}).
\newblock \bibinfo{note}{Accessed: 2024-08-14}.

\bibitem{eloundou2024gpts}
\bibinfo{author}{Eloundou, T.}, \bibinfo{author}{Manning, S.}, \bibinfo{author}{Mishkin, P.} \& \bibinfo{author}{Rock, D.}
\newblock \bibinfo{journal}{\bibinfo{title}{Gpts are gpts: Labor market impact potential of llms}}.
\newblock {\emph{\JournalTitle{Science}}} \textbf{\bibinfo{volume}{384}}, \bibinfo{pages}{1306--1308}, \doiprefix\url{10.1126/science.adj0998} (\bibinfo{year}{2024}).

\bibitem{josten2020robots}
\bibinfo{author}{Josten, C.} \& \bibinfo{author}{Lordan, G.}
\newblock \emph{\bibinfo{title}{Robots at work: Automatable and non-automatable jobs}} (\bibinfo{publisher}{Springer}, \bibinfo{year}{2020}).

\end{thebibliography}

\clearpage

\clearpage
\pagenumbering{arabic} 
\setcounter{page}{1}

\begin{center}
    \LARGE {Supplementary Information for\\}
     \LARGE{``The Potential Impact of Disruptive AI Innovations on U.S. Occupations''}
\end{center}
\vspace{2em}

\setcounter{figure}{0} 
\renewcommand{\thefigure}{S\arabic{figure}} 
\setcounter{table}{0} 
\renewcommand{\thetable}{S\arabic{table}}

\section*{Results of Classifying Patents}

\subsection{Representative Patent Disruption Index Scores and Percentile Rankings}
\label{supsec:reppatents}

Our analysis confirms that patents that previous literature identified as disruptive or consolidating fall within our respective quartile thresholds. 
Specifically, we computed their quartile rankings within our disruption index values finding that disruptive patents consistently fall within the upper quartile, while consolidating patents are concentrated in the lower quartile (Table~\ref{tab:reppat}).

We also manually examined the patents in our dataset that fall within $5\%$ of the thresholds and found that they are correctly classified (see Tables~\ref{tab:patent_disruption_75_80} and \ref{tab:patent_disruption_20_25}).

\begin{table}[htbp]
\centering
\begin{tabular}{p{1.5cm} p{4cm} p{2cm} R{1cm} R{1cm} R{2.3cm} R{2.3cm} p{1.5cm} }
\toprule
\textbf{Patent ID} & \textbf{Patent Title}& \textbf{Assignee}  & \textbf{Year} &\textbf{Disruption Index} & \textbf{Quantile of Disruption Index} & \textbf{Perceived Category} \\
\midrule
US4637464 & In Situ Retorting of Oil Shale with Pulsed Water Purge & Amoco Corp. & 1987  & -0.77  &  0.00 & Consolidating \\
US4573530 & In-Situ Gasification of Tar Sands Utilizinga Combustible Gas &Mobil Oil Corp. & 1988  & -0.83  &  0.00 & Consolidating \\
 US6958436 & Soybean variety SE90346 &MONSANTO TECHNOLOGY, L.L.C.& 2005& -0.89& 0.00& Consolidating\\
  US6063738 & Foamed well cement slurries, additives and methods &Halliburton Energy Services Inc& 2000& -0.07& 0.03 & Consolidating\\
US4658215& Method for induced polarization logging& Shell Oil Co.& 1987 & -0.85& 0.00& Consolidating\\
US4928765 & Method and apparatus for shale gas recovery & Ramex Syn-Fuels International& 1990&-0.69 & 0.00& Consolidating\\
US4724318 & Atomic Force Microscopeand Method for Imaging Surfaces with Atomic Resolution &IBM Corp. & 1988  & 0.13  &  0.92 & Disruptive \\
US5016107 & Electronic still camera utilizing image compression and digital storage & Eastman Kodak Co. & 1991& 0.14& 0.92&  Disruptive \\
US6285999 & Method for node ranking in a linked database&
Leland Stanford Junior University & 1991& 0.44& 0.98  &  Disruptive \\
 US6285999 & Method for node ranking in a linked database&Leland Stanford Junior University
 & 2001& 0.44& 0.98 &  Disruptive \\
  US4399216 & Processes for inserting DNA into eucaryotic cells and for producing proteinaceous materials&Columbia University in the City of New York 
 & 1983& 0.98& 1.00 &  Disruptive \\
US4356429&Organic electroluminescent cell & Eastman Kodak Co.&1982 &0.90&1.00&Disruptive\\
\bottomrule
\end{tabular}
\caption{Illustrative patents from previous literature~\cite{funk2017dynamic}}
\label{tab:reppat}
\end{table}

\begin{table}[htbp]
\centering
\begin{tabular}{p{1.6cm} p{4cm} p{1.5cm} R{1cm} R{1.5cm} p{6cm} }
\toprule
\textbf{Patent ID} & \textbf{Patent Title} & \textbf{Assignee}  & \textbf{Year} &\textbf{Disruption Index}  & \textbf{Analysis} \\
\midrule

US10064688 & Systems and methods for secure transaction management and electronic rights protection & Cilag GmbH International & 2018  & -0.004 &This patent primarily consolidates existing robotic surgical systems by refining how end effectors are controlled. It improves precision and flexibility in surgery but does not introduce a fundamentally new approach to robotic-assisted procedures.  \\
\midrule
US10105136 & Robotically-controlled motorized surgical instrument with an end effector & Cilag GmbH International & 2018 &-0.003 &  This patent advances robotic-assisted surgical instruments by refining motorized control over an end effector's movement. By improving motion conversion mechanisms, it enhances precision and efficiency in surgical procedures. However, rather than introducing a fundamentally new surgical method, it represents a technological refinement of existing robotic surgery tools. \\
\midrule
US10130361 & Robotically-controller motorized surgical tool with an end effector & Cilag GmbH International & 2018  &-0.003 & This patent is primarily consolidating because it enhances existing robotic surgical tools rather than introducing a fundamentally new approach. It refines motorized control, motion conversion, and modular interfacing for greater precision and efficiency in robotic-assisted stapling, but it does not disrupt the established trajectory of robotic surgery. \\
\midrule
US9198714 & Haptic feedback devices for surgical robot & Cilag GmbH International & 2015 & -0.004  & Similar to US9737326, this patent is primarily consolidating, as it enhances existing robotic surgical systems by integrating haptic feedback to improve surgeon control and usability. While it refines human-robot interaction and may expand robotic surgery adoption, it does not fundamentally alter the technological trajectory but rather optimizes and strengthens it. \\
\midrule
US10201365 &Surgeon feedback sensing and display methods & Cilag GmbH International & 2019 & -0.004  & This patent is consolidating, as it enhances existing robotic surgical systems by integrating visual feedback mechanisms to improve surgeon awareness and precision. While it refines feedback display methods, making robotic surgery more intuitive, it does not disrupt or redefine the field but rather optimizes current practices. \\
\bottomrule
\end{tabular}
\caption{Patents in the 20\% to 25\% Disruption Percentile Range}
\label{tab:patent_disruption_20_25}
\end{table}

\begin{table}[htbp]
\centering
\begin{tabular}{p{1.6cm} p{4cm} p{1.5cm} R{1cm} R{1.5cm} p{6cm} }
\toprule
\textbf{Patent ID} & \textbf{Patent Title} & \textbf{Assignee}  & \textbf{Year} &\textbf{Disruption Index}  & \textbf{Analysis} \\
\midrule
US9437186 & Enhanced endpoint detection for speech recognition & Amazon Technologies Inc & 2016 &0.031 & The patent qualifies as disruptive because it challenges the conventional fixed, post-utterance endpointing paradigm. Its two-tiered, dynamic approach to determining speech endpoints can lead to significantly lower latency and more fluid, responsive user interfaces. In doing so, it reshapes the underlying architecture of speech recognition systems, potentially impacting a wide range of applications and setting new industry standards for responsiveness. \\  
 \\
\midrule
US9043018 & Medical device with orientable tip for robotically directed laser cutting and biomaterial application & Intuitive Surgical Operations Inc & 2015 &0.0346&  While the invention consolidates established technology, its disruptive aspects—especially the multifunctional integration that could change surgical workflows and expand procedural possibilities—appear to be more significant. In effect, the device not only enhances the existing system but also has the potential to shift paradigms in minimally invasive robotic surgery. \\
\midrule
US9881616 &Method and systems having improved speech recognition & Qualcomm Inc&2018& 0.026 &The patent is both disruptive and consolidating. It disrupts the status quo by significantly improving speech recognition reliability in challenging environments, while also consolidating a variety of proven techniques into a single, practical solution. \\
\midrule
US9492923 &Generating robotic grasping instructions for inventory items & Amazon Technologies Inc& 2016& 0.034& This patent can be characterized as disruptive in that it introduces an adaptive, learning-based framework for robotic grasping that significantly shifts how robots interact with diverse, unstructured inventory items—paving the way for more flexible and efficient warehouse operations. Concurrently, it is consolidating because it integrates established technologies (sensor fusion, machine learning, human‑robot interaction) into a unified system, thereby refining and extending existing concepts rather than starting from scratch. \\
\midrule
US9284049 & Unmanned aerial vehicle and operations thereof & SZ DJI Technology Co Ltd & 2016  & 0.028& This patent presents a novel design in which interference‑generating electrical components are confined within an internal cavity while interference‑sensitive sensors (such as magnetometers) are positioned externally. At the same time, it integrates and optimizes established principles. As a result, the invention represents a significant evolution—offering meaningful improvements without completely revolutionizing existing technology.  \\
\bottomrule
\end{tabular}
\caption{Patents in the 75\% to 80\% Disruption Percentile Range}
\label{tab:patent_disruption_75_80}
\end{table}

\newpage
\subsection*{Full Rubric for Classifying Tasks based on Characteristics}
\label{sec:full_task_prompts}

The GPT model was provided with a detailed set of instructions to label tasks according to various characteristics. The original prompts are below:

\begin{lstlisting}[language=json, title=Prompt for Classifying Tasks as Interactive or Independent]
{
  "TASK: "{task}" Please label the given task according to the taxonomy below. ## T -- Interactive Label tasks T if the given task is performed in collaboration with others and involves either alignment or co-creation. ## D -- Independent Label tasks D if the given task requires minimal to low levels of coordination with others, even if work product later needs to integrate with work of others. Please write a response in json file format like below: "Task": "Analyzing data","Label of How (T/D)": "D","Explanation": "Data analysis often involves personal tasks like processing datasets, performing statistical tests, and interpreting results, which can be done autonomously if the analyst possesses the necessary skills and information. This independence is facilitated by the nature of the work, which largely involves interacting with data through software and requires concentrated individual effort. Additionally, advancements in data analysis tools and software have made it easier for individuals to manage and analyze large amounts of data efficiently on their own. Thus, while collaboration can enhance aspects of data analysis, especially in complex projects or interdisciplinary fields, much of the analytical work can be effectively conducted independently." "Task": "Investigate and evaluate union complaints or arguments to determine viability","Label of How (T/D)": "T","Explanation": "Investigating and evaluating union complaints or arguments to determine their viability is a task that predominantly requires collaboration and interaction, although it incorporates some independent elements. This role involves engaging with multiple stakeholders such as union representatives, employees, and management to understand each group's perspective. Information gathering might involve independent research, but it frequently necessitates interviews and discussions with involved parties to grasp the full context of each complaint. Additionally, the task often requires coordination with legal advisors and human resources to ensure compliance with legal standards and organizational policies. If mediation is involved, the role distinctly relies on strong interpersonal skills to manage and reconcile differing viewpoints, underscoring the collaborative nature of the task." Once again, please make sure that the response is in json format."

}
\end{lstlisting}

\begin{lstlisting}[language=json, title=Prompt for Classifying Tasks as Variable or Repetitive]
{
 "TASK: "{task}" Please label the given task according to the taxonomy below. ## R -- Repetitive Label tasks R if the task involves performing the same standardized procedures of operations consistently, with little variation over time. ## V -- Variable Label tasks D if the task involves frequent changes in procedures, requiring adaptability and decision-making based on unique circumstances each time. Please write a response in json file format like below: "Task": "assembling components onto a circuit board","Label of Repetitiveness (R/V)": "R", "Explanation": "This task involves placing specific electronic components like resistors, capacitors, and integrated circuits in designated spots on the circuit board and soldering them into place. The task is repeated with each circuit board, following a precise pattern and methodology to ensure consistency and functionality of the final product. Each step is standardized and repeated for multiple units, making the process highly repetitive." "Task": "Investigate and evaluate union complaints or arguments to determine viability","Label of Repetitiveness (R/V)": "V","Explanation": T"he task of investigating and evaluating union complaints or arguments to determine their viability is an example of variable work. This role involves understanding the specific details of each complaint, which can vary widely in nature, context, and seriousness. The process requires analyzing documentation, interviewing involved parties, interpreting labor laws and agreements, and applying these to the unique circumstances of each case. The variability in the tasks arises from the need to adapt approaches based on different legal frameworks, workplace policies, and the specifics of each complaint, necessitating significant human judgment and adaptability." Once again, please make sure that the response is in json format."

}
\end{lstlisting}

\begin{lstlisting}[language=json, title=Prompt for Classifying Tasks as Mental or Physical]
{
"TASK: \"{task}\"\n\nPlease label the given task according to the taxonomy below by choosing one option from the annotations provided.\n\n## P -- Physical\nLabel tasks P if the task involves bodily movement, physical exertion, and the use of physical skills or strength. \n\n## M -- Mental\nLabel tasks M if the task involves cognitive activities that require thinking, problem-solving, decision-making, and the use of intellectual skills. Please write a response in json file format like below:\n\nTask: Building a wooden chair\nLabel of Nature (P/M): P\nExplanation: This task involves physical activities such as cutting, sanding, and assembling pieces of wood using tools. It requires manual labor and physical exertion to shape and join the wood pieces into a finished chair.\n\nTask: Analyzing sales data to identify trends\nLabel of Nature (P/M): M\nExplanation: This task involves cognitive activities such as collecting data, performing statistical analyses, interpreting results, and making decisions based on the findings. It requires intellectual skills and problem-solving abilities to understand and analyze the sales data.\n\nOnce again, please make sure that the response is in json format."
}
\end{lstlisting}

\subsection*{Prolific Survey}

\begin{figure}[h]
    \centering    \includegraphics[width=0.8\linewidth]{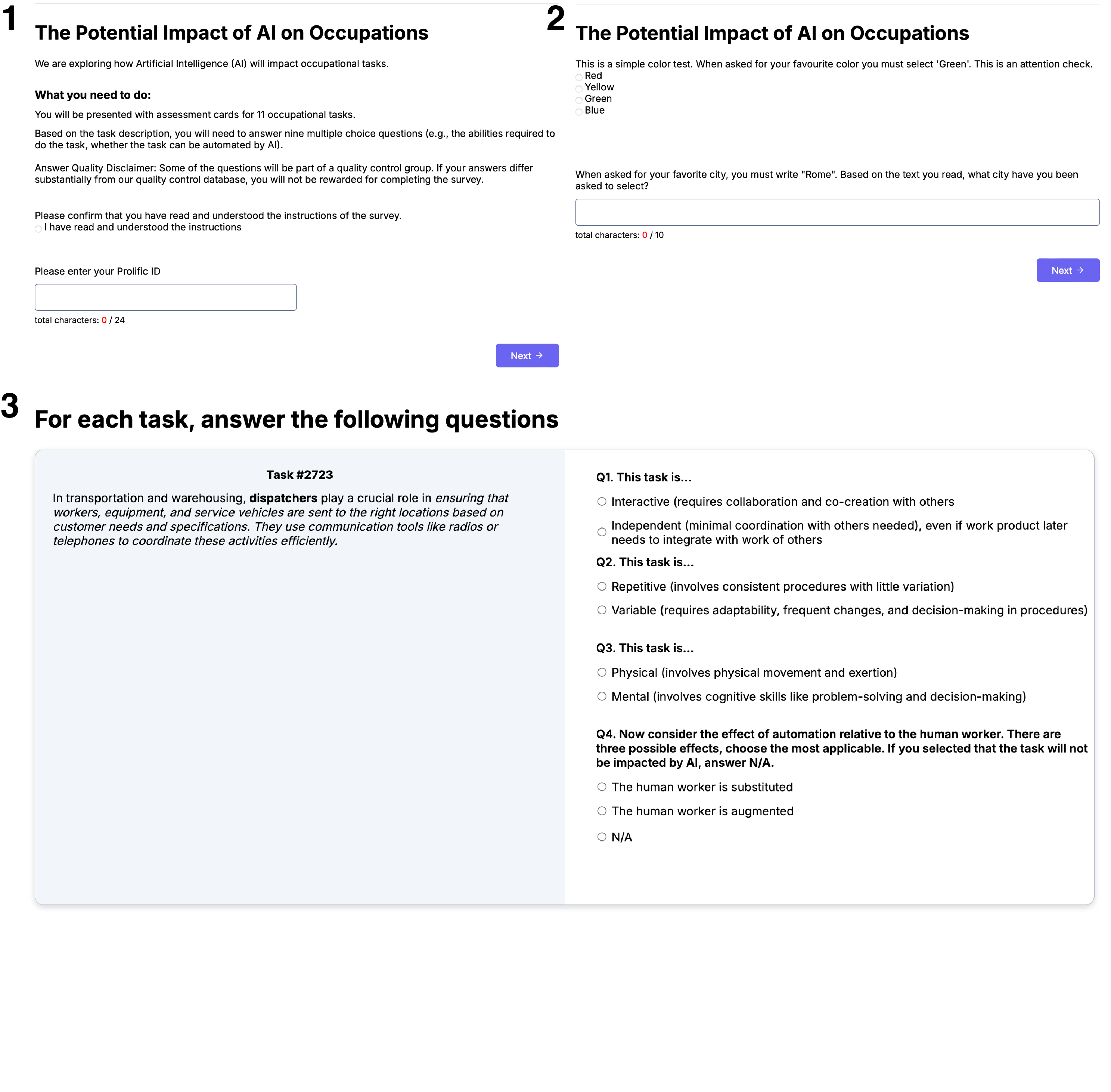}
    \caption{\textbf{Survey Interface Used in the Prolific Study.} 1. Participants were first presented with instructions and definitions to guide them through the classification process. 2. An attention check followed to ensure the reliability of responses. 3. Participants were then asked to classify occupational tasks across multiple dimensions, including time sensitivity, variability, interactivity, and physicality, among others. To help participants understand each task, we provided the corresponding occupation and industry information.}
    \label{fig:survey}
\end{figure}

The survey conducted on Prolific was designed to collect human annotations on a subset of occupational tasks (Figure~\ref{fig:survey}). Participants were presented with tasks and asked to classify them according to the same rubric used by GPT model. 

The survey included multiple-choice questions that mirrored the categories provided to GPT model.

\subsection{Task Network Construction and Community Detection}
\label{sub:network_louvain}
\begin{figure}[h]
    \centering    
    \includegraphics[width=0.95\linewidth]{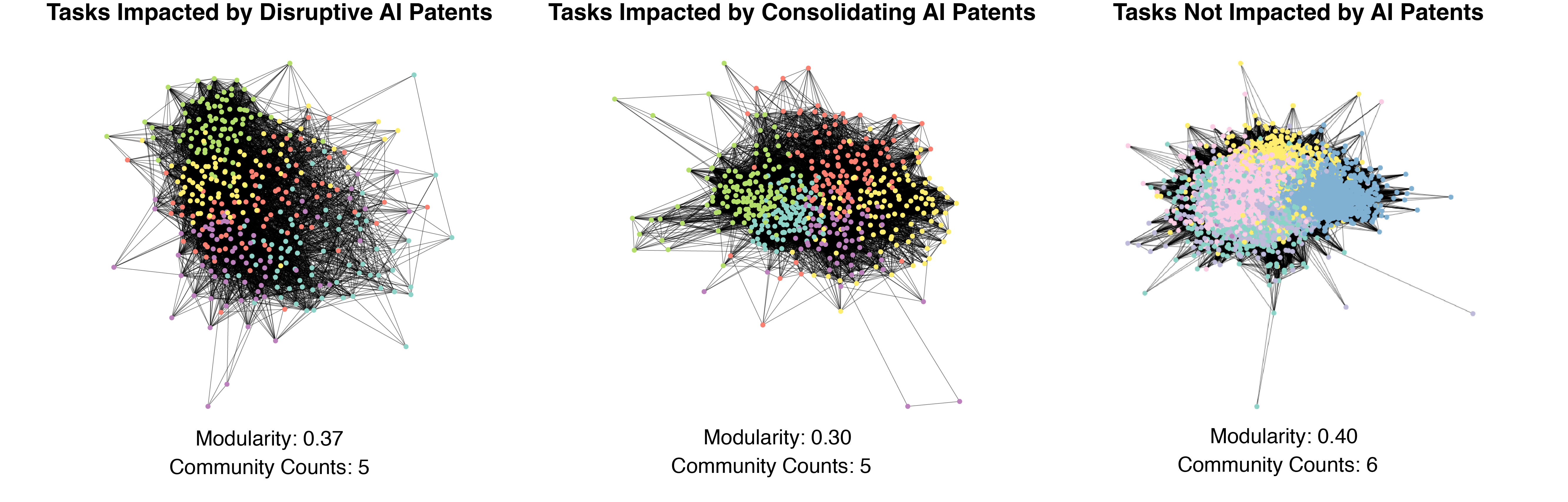}
    \caption{\textbf{Network Visualization of Tasks Impacted by Disruptive and Consolidating AI Patents, and Tasks Not Impacted by AI Patents.} Each network represents tasks as nodes, with connections between nodes established based on cosine similarity, specifically when the similarity is in the top 95\textsuperscript{th} percentile among all task pairs. The Louvain method was applied to cluster the networks, resulting in distinct communities.}
    \label{fig:louvain}
\end{figure}

To identify representative tasks impacted by disruptive and consolidating AI patents, as well as tasks not impacted by AI patents, we constructed a network for each set of tasks. In these networks, each task is depicted as a node, with connections established between nodes when their cosine similarity ranks in the top 95\textsuperscript{th} percentile among all task pairs. We then applied the Louvain method to cluster the networks and calculated their modularity scores. The network for tasks impacted by disruptive AI patents exhibited a modularity score of 0.37, revealing 5 distinct communities, while the network for tasks affected by consolidating AI patents showed a modularity of 0.3 with 5 communities. In contrast, the network for tasks not impacted by AI patents had a higher modularity score of 0.4 and formed 6 communities. The Louvain method detects communities by optimizing modularity, a metric that evaluates how densely connected the nodes are within clusters relative to connections between different clusters~\cite{blondel2008fast}. A network with perfect modularity would score 1, indicating very strong community structure, whereas a network lacking any community structure would score -0.5. The resulting network is illustrated in Figure~\ref{fig:louvain}, with each community represented by a distinct color.

\begin{figure}[t]
    \centering
    \includegraphics[width=0.9\linewidth]{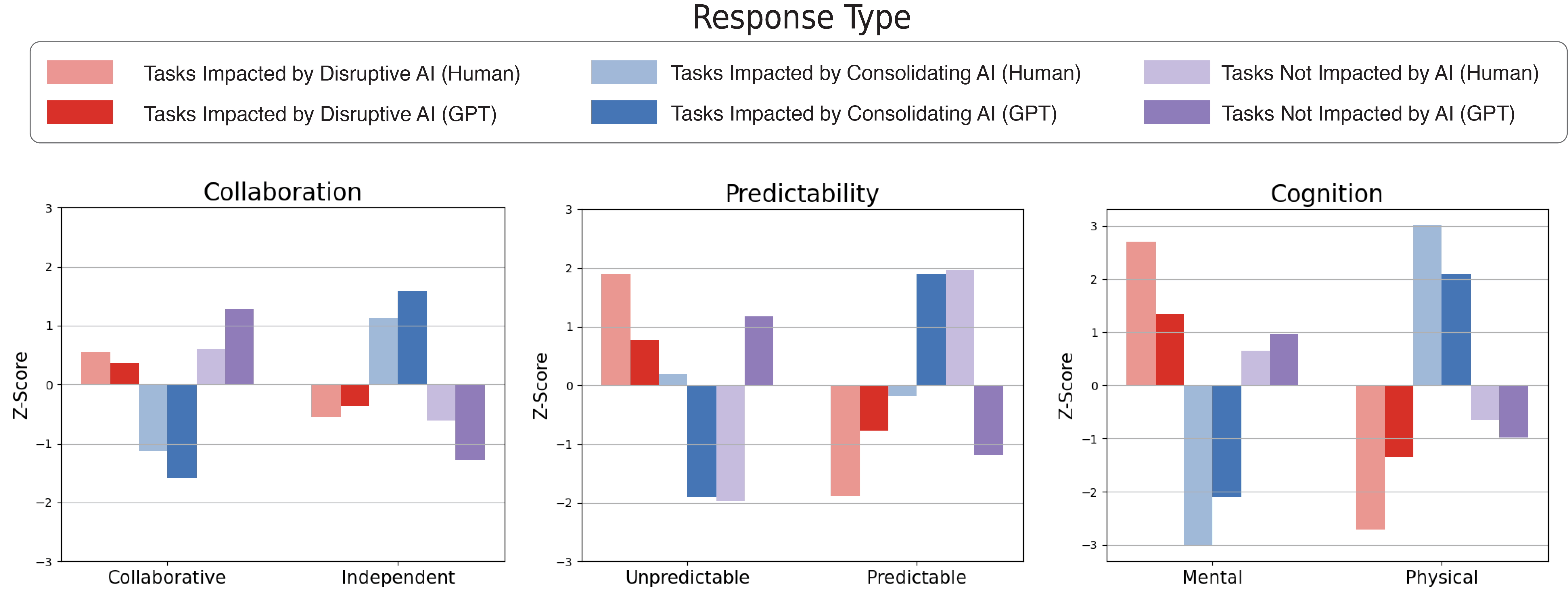}
    \caption{\textbf{$z$-Scores for Task Characteristics Based on Original GPT and Human Annotations.} }
    \label{fig:original_response}
\end{figure}

\subsection{Comparison of Task Classification Results Between GPT and Human Annotators}
\label{sec:original_gpt_human}

In this section, we provide a detailed comparison between the task classifications conducted by GPT-4o and human annotators using representative samples. These representative samples were selected from each cluster in the networks constructed in the previous section (see \nameref{sub:network_louvain}). Within each cluster, tasks were chosen for annotation proportionally to the cluster size. The selection process prioritized nodes with the highest degrees, followed by the next highest, ensuring that the most central tasks in each cluster were included. The total number of representative samples was set to 5\% of the total tasks impacted by disruptive and consolidating AI, resulting in 20 tasks impacted by disruptive AI and 28 tasks impacted by consolidating AI. For tasks not impacted by AI, due to the large size of the original task set, we selected a similar number of representative tasks, which is in total 29 tasks.

The $z$-scores calculated from the original responses of human annotators and GPT model are shown in Figure~\ref{fig:original_response}. These scores indicate how likely tasks are to exhibit specific characteristics compared to a null model, which was generated by randomizing the characteristics of tasks while maintaining the overall distribution of answers.

In both human and GPT annotations, tasks impacted by disruptive AI patents were more likely to be characterized as collaborative, unpredictable, and mental than tasks impacted by consolidating AI patents. The $z$-score for tasks classified as collaborative by disruptive AI patents were 0.548 for humans and 0.367 for GPT, while the $z$-score for unpredictable tasks is 1.891 for human and 0.765 for GPT. For tasks classified as mental, the $z$-scores were 2.706 for humans and 1.347 for GPT. Conversely, tasks impacted by consolidating AI patents were more frequently associated with being independent and physical. The $z$-scores for tasks classified as independent by consolidating AI patents were 1.129 for humans and 1.590 for GPT, and for physical tasks, the $z$-scores were 3.012 for humans and 2.090 for GPT. The $z$-score of predictable tasks annotated by GPT was 1.890. While the $z$-score for predictable tasks classified by human is slightly below zero, it is still notably higher than the score for tasks associated with disruptive AI patents. This suggests that while both disruptive and consolidating AI patents lead to a higher likelihood of tasks being classified as unpredictable, disruptive AI patents show a much stronger tendency to be unpredictable than consolidating AI patents. 

However, in the classification pattern of tasks not impacted by AI, a discrepancy emerged between human annotators and GPT. While both groups agreed that these tasks are more likely to be collaborative with a $z$-score of 0.606 for human response and 1.283 for GPT, and to be mental with $z$-scores of 0.650 and 0.976, respectively, their classification diverged on other aspects. GPT classified these tasks as more likely to be unpredictable  with $z$-score 1.178. In contrast, human annotators indicated a tendency towards classifying these tasks as predictable with $z$-scores of 1.965.

Interestingly, GPT did not classify any tasks under ``Not Impacted by AI'' while human annotators did use these categories for some tasks. 

Furthermore, the $z$-scores for tasks classified as Not Impacted by AI in the consolidating category are relatively low but positive at 0.256 for humans, while for the disruptive category, the $z$-score is negative with -0.375. The $z$-scores for tasks classified as Not Impacted by AI in the Not impacted category was also positive with 0.0937.

\subsection*{High Agreement Between GPT-4o's Task Labeling and Human Annotations} 
\label{sec:validation}
To validate the performance of GPT-4o in classifying task characteristics, we conducted a comparison between its annotations and those provided by 24 human annotators for a subset of tasks. These subset of tasks were selected based on their centrality within different clusters of the entire task set, ensuring that they reflect the key characteristics and variations across the broader dataset. Specifically, we sampled 5\% of the representative tasks impacted by disruptive and consolidating AI, resulting in 20 tasks influenced by disruptive AI and 28 by consolidating AI. For tasks not impacted by AI, we selected a similar number of representative tasks, totaling 29, due to the large size of the initial task set. We observed moderate to high agreement rates for task characteristics; these agreement rates are broadly consistent with those reported in previous studies~\cite{eloundou2024gpts}. Specifically, the agreement rate for unpredictable \emph{vs.} predictable was 66.24\%, for mental  \emph{vs.} physical was 72.73\%, and for collaborative  \emph{vs.} individual was 79.22\%. Furthermore, after resolving the discrepancies between the GPT-4o model's and human annotations through a consensus process involving three authors, we found that the authors' final annotations for the tasks with discrepancies were more closely aligned with GPT-4o than with the original human annotations. The alignment rate was 52.38\% for mental \emph{vs.} physical classification, 81.81\% for unpredictable \emph{vs.} predictable classification, and 81.25\% for collaborative \emph{vs.} individual classification. \smallskip

Since we checked and confirmed that GPT-4o's annotations are accurate, our results include all the tasks impacted by disruptive AI (397 tasks) and consolidating AI (536 tasks), along with a balanced and fair sample of 933 tasks that are not impacted by AI.

\section*{Results of Task Classification}\setcurrentname{Results of Task Classification}\label{sec:app_results}\addcontentsline{toc}{section}{Results of Task Classification}\phantomsection

\subsection*{Literature Review on Task Characteristics and Automation} 
\label{supsec:literature}

\revision{To identify relevant literature on job and task characteristics in the context of technological change, we developed a structured search strategy. We first selected seven foundational papers that extensively examine the relationship between job or task characteristics and technological change \cite{autor2003skill, josten2020robots, autor2013growth, autor2015untangling, frey2017future, felten2021occupational, webb2019impact}.}

{
\sloppy
Based on the titles of these references, we developed keyword combinations to search the titles of the relevant studies. The titles of these references fall into two distinct categories. one group of studies hypothesizes how job roles, skills, and task requirements evolve in response to technological change---analyzed through shifts in labor market dynamics (e.g., \cite{autor2003skill, autor2013growth, autor2015untangling})---and another group explicitly examines the impact of technological change on jobs, skills, and tasks (e.g., \cite{josten2022automation, felten2021occupational, frey2017future, webb2019impact}). To capture the first category, we incorporated search terms such as \texttt{"skill*"} or \texttt{"task*"} in combination with \texttt{"labor market"} or \texttt{"labour market"}. For the second category, our query expanded to include terms like \texttt{"work*"}, \texttt{"occupation*"}, \texttt{"employment"}, \texttt{"job*"}, or \texttt{"skill*"} along with \texttt{"automation"}, \texttt{"computerisation*"}, \texttt{"exposure to artificial intelligence"}, \texttt{"technological change"}, or \texttt{"impact of artificial intelligence"}.

To refine the search and ensure relevance, we excluded keywords that were either too narrow or unrelated to our research focus. Specifically, we removed industry- or job-specific terms such as \texttt{``mining''}, 
\texttt{``medical*''}, \texttt{``education''}, \texttt{``healthcare*''}, \texttt{``manufact*''}, \texttt{``housework''},\texttt{``arts''},  and \texttt{``salesperson*''} to prevent overrepresentation of sector-specific studies. We also excluded studies focusing on worker well-being, satisfaction, stress, and health rather than the intrinsic nature of job tasks by omitting terms such as \texttt{``stress''}, \texttt{``depression''}, \texttt{``attitude''}, \texttt{``satisfaction''}, \texttt{``environment*''}, \texttt{``sustainability''},  \texttt{``union''}, \texttt{``COVID''}, and \texttt{``health''}. Additionally, to maintain focus on job characteristics rather than regulatory or legal aspects, we excluded \texttt{``policy''}, \texttt{``legal''}, \texttt{``law''}, and \texttt{``legislation''}. Finally, we removed unrelated conceptual studies by omitting terms like \texttt{``framework''}, and \texttt{``model''}.}, the studies focusing on retrospective narratives rather than technological impacts by removing the terms \texttt{``history''}, and the title with the unrelated terms including ``work'' such as \texttt{``workplace''},
\texttt{``workshop''}, \texttt{``workflow''}.

\revision{The Web of Science query used for the literature search was as follows:}
\lstset{basicstyle=\ttfamily\small, breaklines=true}
\begin{lstlisting}
TI = (   
    ( ( ("skill*" OR "task*" ) AND ("labor market" OR "labour market")) 
      OR 
      ( ("work*" OR "occupation*" OR "employment" OR "job*" OR "skill*" OR "labor market" OR "labour market") 
         AND 
        ("automation" OR "computerisation*" OR "exposure to artificial intelligence"  OR "technological change" OR "impact of artificial intelligence") 
      ) 
    ) 
) 
NOT TI = ("policy" OR "policies" OR "legal" OR "law" OR "legislation" OR "workflow" OR "work flow" OR "workshop" OR "education" OR "educational"  OR "history" OR "histories" OR "medical*" OR "healthcare*" OR "training" OR "agricultur*" OR "mining" OR "sustainability" OR "environment*" OR "arts" OR "manufact*" OR "salesperson*" OR "model" OR "framework" OR "depression" OR "stress" OR "COVID" OR "health" OR "housework" OR "graduate*"  OR "attitude*" OR "union*" OR "satisfaction" OR "workplace safety") 
\end{lstlisting}

\revision{Using these keywords, we searched the Web of Science database, restricting results to English-language articles classified under Economics, Business, or Multidisciplinary Science, with a citation topic of Economics. Applying these search criteria, we initially retrieved 181 papers from Web of Science. }

\section*{Geographic Distribution of Disruptive AI Patent and Consolidating AI Patent Assignee}
In the main results, we identified a regional pattern in AI patent's impact on tasks by contrasting the geographic distribution of disruptive versus consolidating AI patents across U.S. states. To explore this further, we compared the geographical distribution of the assignees of disruptive and consolidating AI patents to the distribution of general AI patents. Specifically, we calculated this difference by subtracting the state's proportion of general AI patents from its proportion of disruptive or consolidating AI patents.

\begin{figure}[t]
    \centering
    \includegraphics[width=0.9\linewidth]{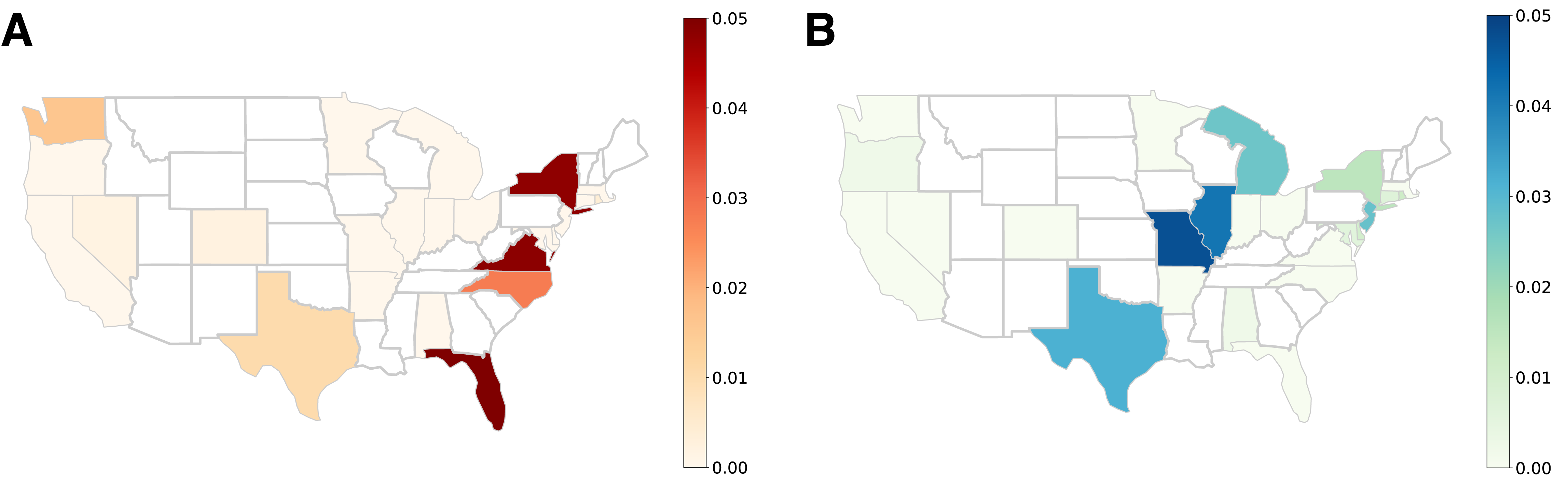}
    \caption{\textbf{Geographic Distribution of Disruptive and Consolidating AI Patents in the U.S. Relative to General AI Patents.}  A) Distribution of disruptive AI patents across U.S. states, with color intensity indicating regions where these patents are more concentrated relative to the overall AI patent distribution impacting tasks. Regions with white color indicates that the state has no disruptive AI patents impacting occupational tasks. B) Distribution of consolidating AI patents, with color intensity representing regions where these patents are more or less concentrated relative to the overall AI patent distribution. Regions with white color indicates that the state has no consolidating AI patents impacting occupational tasks.}
    \label{fig:geometry}
\end{figure}

The findings are visualized in Figure.~\ref{fig:geometry}. The results still reveal a significant difference between the distribution of consolidating and disruptive AI patents compared to the general AI patents. Disruptive AI patents are more concentrated on the East Coast, while consolidating AI patents show a different regional pattern compared to general AI patents.

\section*{Impact of Disruptive and Consolidating AI on Industry}

\begin{figure}
    \centering
    \includegraphics[width=0.5\linewidth]{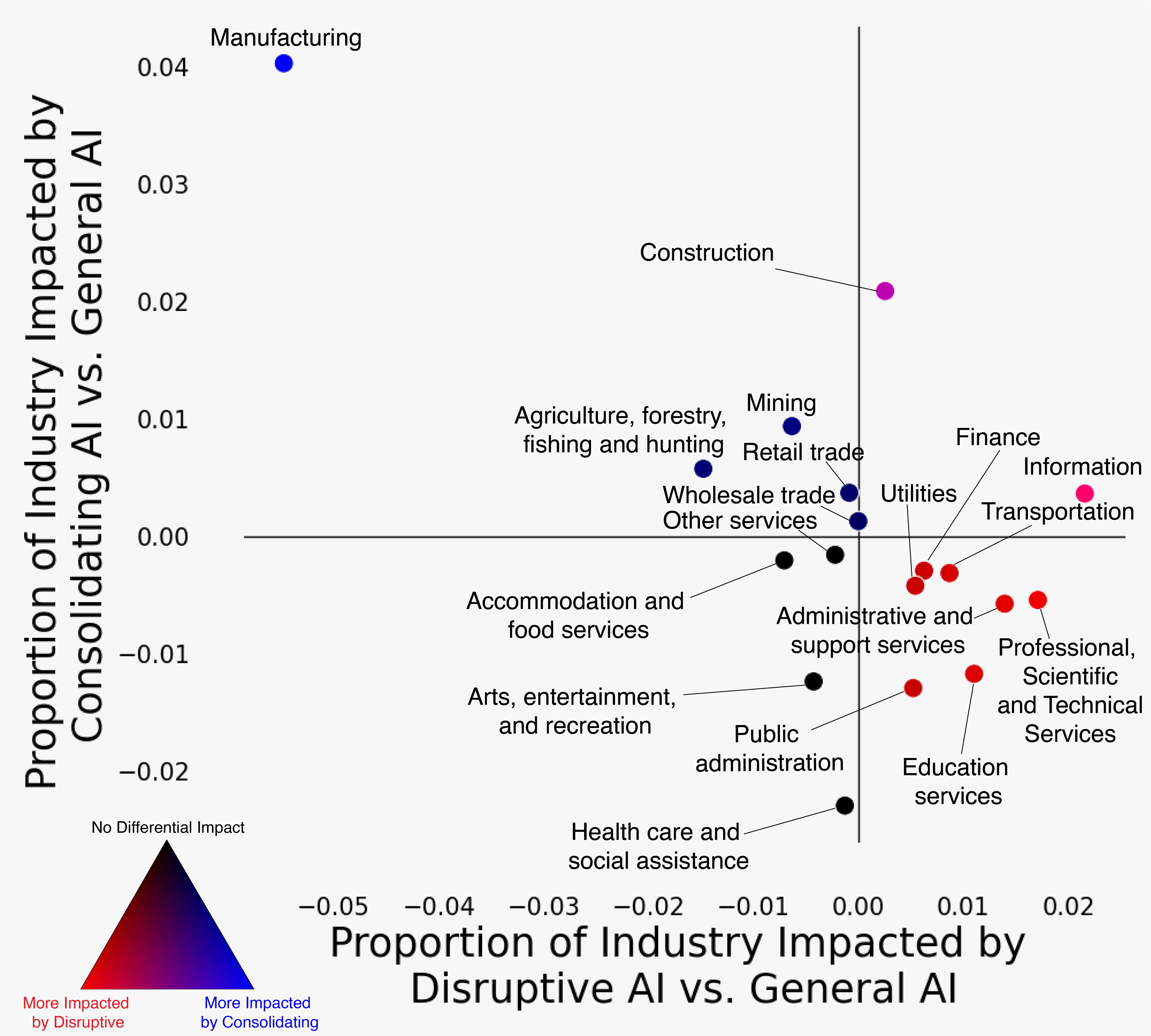}
    \caption{Scatter plot showing the proportion of tasks in each industry sector impacted by disruptive and consolidating AI patents relative to all AI patents. The X-axis represents the difference between the share of tasks impacted by disruptive AI patents and that impacted by all AI patents, while the Y-axis represents the corresponding difference for consolidating AI. For instance, an X-axis value of 0.01 indicates that the share of tasks impacted by disruptive AI patents is 1\% higher than that impacted by general AI patents.}
    \label{fig:all_industry}
\end{figure}

In the main text, we presented the top five industry sectors most affected by disruptive and consolidating AI patents relative to all AI patents (Fig.~\ref{fig:bubble_industry}). In Fig.~\ref{fig:all_industry}, we extend this to every industry sector by displaying the differences between the share of tasks impacted by disruptive AI patents and those impacted by all AI patents, alongside the corresponding differences for consolidating AI in a scatter plot. As noted earlier, traditional and production-oriented sectors tend to be more impacted by consolidating AI, whereas service and technology sectors are more influenced by disruptive AI.  \revision{Among these, the Information sector stands out with a statistically significant increase in exposure to disruptive AI relative to overall AI exposure (p = 0.033), suggesting it is disproportionately impacted by disruptive innovations. In contrast, Manufacturing and Construction exhibit significantly higher exposure to consolidating AI (p = 0.027 and p = 0.048, respectively), reinforcing their alignment with incremental technological improvements. P-values are derived from two-proportion z-tests that compare the proportion of tasks linked to disruptive or consolidating AI patents with the overall proportion of AI-exposed tasks within each industry. The full set of results is available in Tables~\ref{tab:diff_sorted} and~\ref{tab:consolidating_diff_sorted}.}

\begin{table}[ht]
\centering
\begin{tabular}{llrrr}
\hline
\textbf{Industry} & \textbf{Prop. Disruptive} & \textbf{Prop. Overall AI} & \textbf{Z-Stat} & \textbf{P-Value} \\
\hline
\textbf{Information} & \textbf{0.0302} & \textbf{0.0149} & \textbf{2.129} & \textbf{0.033} \\
Administrative and support and waste management & 0.0176 & 0.0077 & 1.873 & 0.061 \\
Professional, scientific and technical Services & 0.0630 & 0.0513 & 0.946 & 0.344 \\
Finance and insurance & 0.0101 & 0.0056 & 1.011 & 0.312 \\
Education services & 0.0302 & 0.0226 & 0.912 & 0.362 \\
Transportation and warehousing & 0.0605 & 0.0549 & 0.442 & 0.659 \\
Utilities & 0.0277 & 0.0241 & 0.422 & 0.673 \\
Public administration & 0.0529 & 0.0497 & 0.262 & 0.793 \\
Other services, except public administration & 0.0151 & 0.0169 & -0.257 & 0.797 \\
Retail trade & 0.0227 & 0.0236 & -0.110 & 0.912 \\
Construction & 0.0353 & 0.0338 & 0.142 & 0.887 \\
Health care and social assistance & 0.1662 & 0.1692 & -0.145 & 0.885 \\
Arts, entertainment, and recreation & 0.0076 & 0.0108 & -0.580 & 0.562 \\
Wholesale trade & 0.0025 & 0.0026 & -0.016 & 0.987 \\
Accommodation and food services & 0.0050 & 0.0103 & -0.984 & 0.325 \\
Mining & 0.0025 & 0.0072 & -1.062 & 0.288 \\
Agriculture, forestry, fishing and hunting & 0.0025 & 0.0133 & -1.842 & 0.066 \\
Manufacturing & 0.1688 & 0.2108 & -1.894 & 0.058 \\
\hline
\end{tabular}
\caption{Difference Between Disruptive AI and Overall AI Exposure by Industry.}
\label{tab:diff_sorted}
\end{table}

\begin{table}[ht]
\centering
\begin{tabular}{llrrr}
\hline
\textbf{Industry} & \textbf{Prop. Consolidating} & \textbf{Prop. Overall AI} & \textbf{Z-Stat} & \textbf{P-Value} \\
\hline
\textbf{Manufacturing} & \textbf{0.0448} & \textbf{0.2108} & \textbf{2.217} & \textbf{0.0266} \\
\textbf{Construction} & \textbf{0.0184} & \textbf{0.0338} & \textbf{1.977} & \textbf{0.0480} \\
Mining & 0.0077 & 0.0072 & 1.696 & 0.0899 \\
Retail trade & 0.0044 & 0.0236 & 0.583 & 0.5602 \\
Information & 0.0038 & 0.0149 & 0.625 & 0.5323 \\
Agriculture, forestry, fishing and hunting & 0.0053 & 0.0133 & 0.914 & 0.3609 \\
Wholesale trade & 0.0012 & 0.0026 & 0.452 & 0.6515 \\
Transportation and warehousing & 0.0011 & 0.0549 & 0.099 & 0.9214 \\
Other services, except public administration & -0.0001 & 0.0169 & -0.021 & 0.9832 \\
Professional, scientific and technical Services & -0.0009 & 0.0513 & -0.085 & 0.9325 \\
Accommodation and food services & -0.0009 & 0.0103 & -0.191 & 0.8487 \\
Finance and insurance & -0.0019 & 0.0056 & -0.543 & 0.5872 \\
Administrative and support and waste management & -0.0040 & 0.0077 & -0.986 & 0.3244 \\
Public administration & -0.0068 & 0.0497 & -0.654 & 0.5133 \\
Health care and social assistance & -0.0069 & 0.1692 & -0.380 & 0.7042 \\
Education services & -0.0076 & 0.0226 & -1.094 & 0.2737 \\
Arts, entertainment, and recreation & -0.0089 & 0.0108 & -1.949 & 0.0513 \\
Utilities & -0.0017 & 0.0241 & -0.231 & 0.8174 \\
\hline
\end{tabular}
\caption{Difference Between Consolidating AI and Overall AI Exposure by Industry. }
\label{tab:consolidating_diff_sorted}
\end{table}

\section*{Geographic Distribution of Disruptive AI Patent and Consolidating AI Patent Assignee by Industry}
We further investigated how the distributions of disruptive and consolidating AI patents with impact on occupational tasks vary across different industry sectors. Specifically, we focused on the top four industries with the highest number of AI patents impacting occupational tasks: manufacturing, healthcare, transportation, and scientific and technical services. By comparing the distribution of disruptive and consolidating AI patents within these industries to the general AI patent landscape, we aimed to uncover industry-specific geographic patterns.

\begin{figure}[t]
    \centering
    \includegraphics[width=0.96\linewidth]{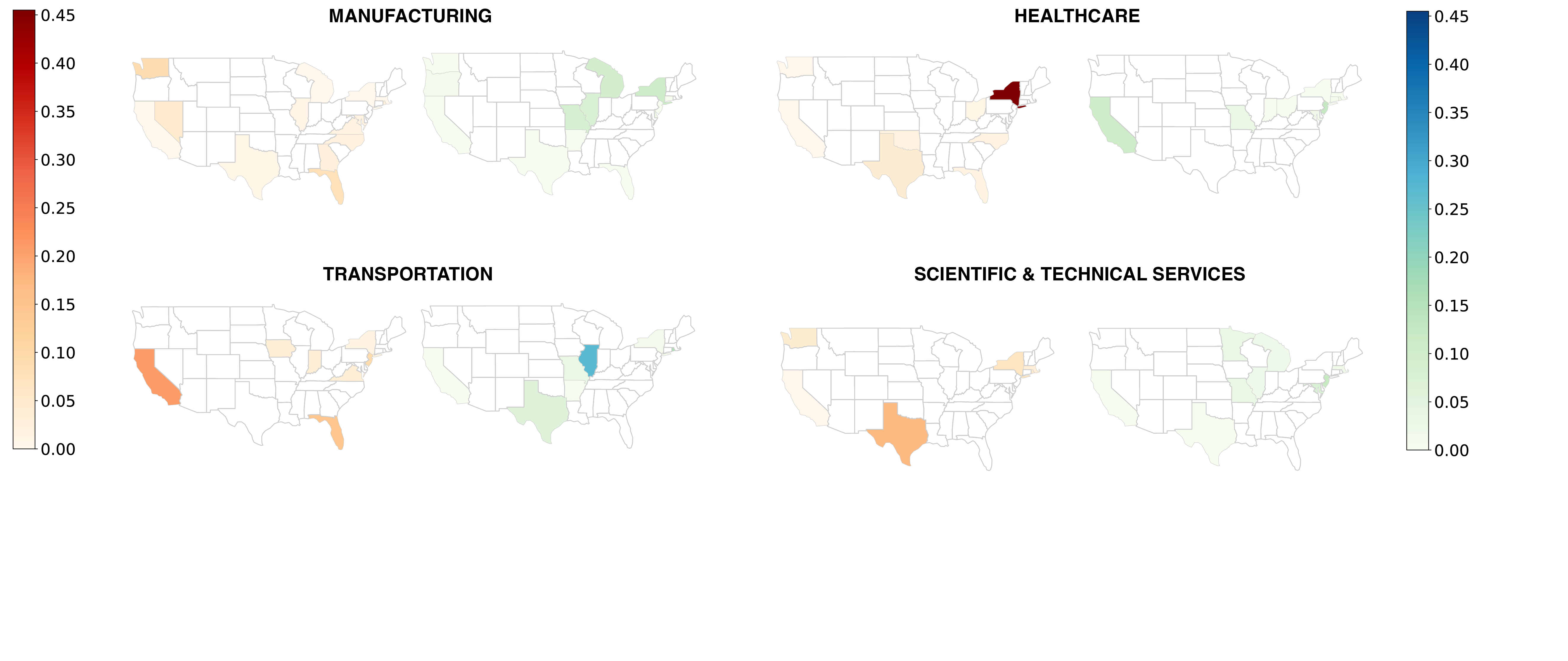}
    \caption{\textbf{Geographic Distribution of Disruptive and Consolidating AI Patents in the U.S. Relative to General AI Patents.} Distribution of disruptive AI patents and consolidating AI patents having an impact on different industries across U.S. states. Color intensity indicating regions where these patents are more concentrated relative to the overall AI patent distribution impacting tasks. Regions with white color indicates that the state has no disruptive AI patents or consolidating AI patents impacting occupational tasks.}
    \label{fig:geometry_industry}
\end{figure}

Interestingly, for manufacturing and transportation, disruptive AI patents are predominantly concentrated in coastal areas, aligning with the overall pattern observed when analyzing all disruptive AI patents together (Figure~\ref{fig:geometry_industry}). For the healthcare, disruptive AI patents in this industry are highly concentrated in New York, while consolidating patents are more in California. For scientific and technical services, both disruptive and consolidating AI patents are primarily produced in similar regions, including California and Texas, with a higher concentration of disruptive AI patents in Texas.

\revision{\section*{Relationship between industry-level job vacancy rates and impact of disruptive and consolidating AI across all industry sectors.}}

\begin{figure}[t]
    \centering
    \includegraphics[width=0.96\linewidth]{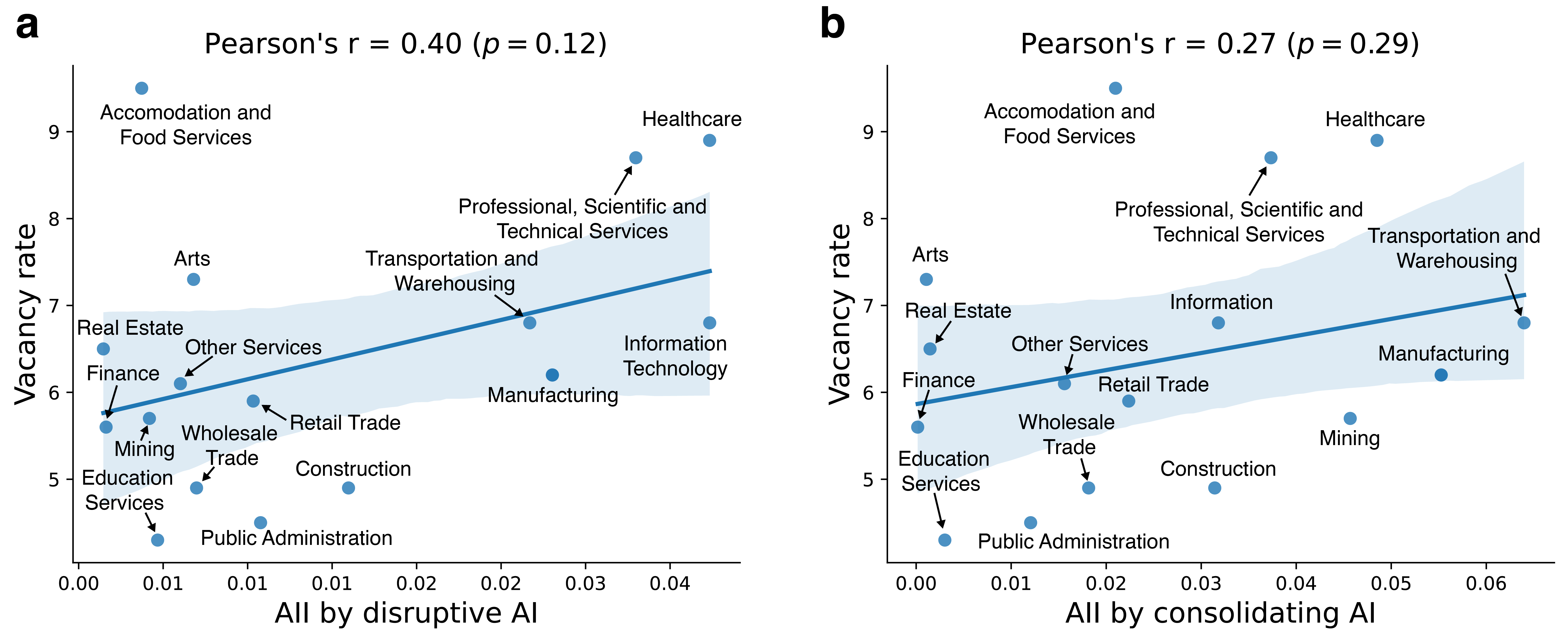}
    \caption{Relationship between job vacancy rates and sector-level impact of disruptive and consolidating AI, shown across all industry sectors.}
    \label{fig:vacancy_rate_si}
\end{figure}

\noindent\revision{Even when including the outlier sector (e.g., accommodation and food services), the correlation between job vacancy rates and sector-level impact of disruptive AI remains stronger than that for consolidating AI (Figure~\ref{fig:vacancy_rate_si}). Both types of AI impact still show a positive relationship with vacancy rates. In particular, the correlation between disruptive AI impact and vacancy rates yields a Pearson's correlation coefficient of $r = 0.40 (p = 0.12)$, while the correlation for consolidating AI is weaker, at $r = 0.27 (p = 0.29)$, though both results are not statistically significant.}

\end{document}